\documentclass[12pt]{article}
\usepackage{amsmath, amssymb}
\usepackage{amsfonts}
\usepackage{graphicx,psfrag,epsf}
\usepackage{enumerate}
\usepackage{natbib, xcolor}
\usepackage{url} 
\usepackage[noend]{algpseudocode}
\usepackage{algorithm}
\usepackage{booktabs}
\usepackage{multirow}
\usepackage{verbatim}
\usepackage{array}
\usepackage{caption}
\usepackage{setspace}
\usepackage{authblk}
\usepackage[compact]{titlesec}

\newcommand{\ra}[1]{\renewcommand{\arraystretch}{#1}}
\newcommand{\blind}{0}
\newcommand{\boldeta}{\mbox{\boldmath $\eta$}}
\newcommand{\bbeta}{\mbox{\boldmath $\beta$}}
\newcommand{\btheta}{\mbox{\boldmath $\theta$}}

\newcommand{\bgamma}{\mbox{\boldmath $\gamma$}}

\newcolumntype{L}[1]{>{\raggedright\let\newline\\\arraybackslash\hspace{0pt}}m{#1}}
\newcolumntype{C}[1]{>{\centering\let\newline\\\arraybackslash\hspace{0pt}}m{#1}}
\newcolumntype{R}[1]{>{\raggedleft\let\newline\\\arraybackslash\hspace{0pt}}m{#1}}
\algnewcommand\And{\textbf{and}}
\DeclareMathOperator*{\argmax}{arg\,max}
\DeclareMathOperator*{\argmin}{arg\,min}
\newtheorem{theorem}{Theorem}
\newtheorem{proposition}{Proposition}

\begin{document}

\def\spacingset#1{\renewcommand{\baselinestretch}%
{#1}\small\normalsize} \spacingset{1}


\if0\blind
{

  \title{\bf Parameter-Expanded ECME Algorithms for \\[2mm] Logistic and Penalized Logistic Regression}
\author[1]{Nicholas C. Henderson}
\author[1]{Zhongzhe Ouyang}
\affil[1]{{\small Department of Biostatistics, University of Michigan, Ann Arbor}}

\date{}
  \maketitle
} \fi

\if1\blind
{
  \bigskip
  \bigskip
  \bigskip
  \begin{center}
    {\LARGE\bf Parameter-Expanded ECME Algorithms for \\[2mm] Logistic and Penalized Logistic Regression}
\end{center}
  \medskip
} \fi

\bigskip
\begin{abstract}
Parameter estimation in logistic regression is a well-studied
problem with the Newton-Raphson method being one of the most prominent
optimization techniques used in practice. A number of monotone optimization methods including minorization-maximization (MM) algorithms, expectation-maximization (EM) algorithms and related variational Bayes approaches offer a family of useful alternatives guaranteed to increase the logistic regression likelihood at every iteration. In this article, we propose a modified version of a logistic regression EM algorithm which can substantially improve computationally efficiency while preserving the monotonicity of EM
and the simplicity of the EM parameter updates. By introducing an additional latent parameter and selecting this parameter to maximize the penalized observed-data log-likelihood at every iteration, our iterative algorithm can be interpreted as a parameter-expanded expectation-condition maximization either (ECME) algorithm, and we demonstrate
how to use the parameter-expanded ECME with an arbitrary choice of weights and penalty function.
In addition, we describe a generalized version of our parameter-expanded ECME algorithm that can be tailored to the challenges encountered in specific high-dimensional problems, and we study several interesting connections between this generalized algorithm and other well-known methods. 
Performance comparisons between our method, the EM algorithm,
and several other optimization methods are presented using 
a series of simulation studies based upon both real and synthetic datasets.
\end{abstract}

\noindent%
\vfill

\newpage
\spacingset{1.45}

\section{Introduction}
\label{sec:intro}

Logistic regression is one of the most well-known statistical 
methods for modeling the relationship between a binomial
outcome and a vector of covariates. Popular iterative
method for optimizing the logistic regression objective
function include the Newton-Raphson/Fisher scoring method and
gradient descent methods. While typically converging very rapidly,
Newton-Raphson is not guaranteed to converge
without additional algorithm safeguards imposed such as step-halving (\cite{marschner2011}, \cite{Lange2012}), 
and the behavior of Newton-Raphson can be very unstable for
many starting values. Because of this, more stable, monotone algorithms which are guaranteed to improve the value of the objective function at each iteration can provide a useful alternative.
While gradient descent and proximal gradient descent methods have seen
a recent resurgence due to their scalability
in problems with a large number of parameters, the
convergence of gradient descent can be extremely sluggish, particularly if one does not use an adaptive steplength finding procedure.

In this article, we propose and explore a new monotone
iterative method for parameter estimation in logistic regression.
Our method builds upon an EM algorithm derived from the P\'olya-Gamma latent variable representation of the distribution of a binomial random variable with a logistic link function \citep{Polson2013, Scott2013}. Specifically, we use a parameter-expanded version of the P\'olya-Gamma representation
to build a parameter-expanded expectation-conditional maximization either 
(PX-ECME) algorithm \citep{Lewandowski2010} for maximizing the weighted log-likelihood of interest. Our approach combines the parameter-expanded EM framework \citep{Liu1998} and conditional maximization versions of the EM algorithm described in, for example, \cite{Meng1993} and \cite{Liu1994}. 
More specifically, we perform two conditional maximization steps where the first step performs an ``M-step" with a fixed value of a latent scale parameter and the second step finds the value of the latent scale parameter maximizing the observed-data log-likelihood. Combining these steps leads to a direct parameter-updating scheme that is essentially  
a scalar multiple of the EM parameter update in each iteration.

While Newton-Raphson and gradient descent procedures are highly effective and will continue to be widely used computational techniques for logistic regression, 
EM-type algorithms for logistic regression and accelerated versions thereof
have a number of distinct advantages, and hence, these algorithms merit further study and development.
First, a major benefit of EM-type algorithms is that each iteration increases the likelihood which
results in a very stable, directly implementable algorithm that
does not require one to implement additional safeguards or step
length finding procedures. Another main advantage of EM-type procedures
is that they often work well in more complex models where one has additional latent variables within a larger, overall probability model. 
For example, a common use of the EM algorithm is in settings where one has missing covariate values (e.g., \cite{ibrahim1990, ibrahim1999}). In such cases, one can often obtain a straightforward EM update of the regression parameters of interest whereas the Newton-Raphson method or other procedures can have very different forms with potentially unknown performance.

While EM algorithms have several notable advantages, slow
convergence of EM is a common issue that often hampers their more widespread adoption. Fortunately, there are numerous techniques which 
can accelerate the convergence of EM while maintaining both the 
monotonicity property of EM and the simplicity of the original EM parameter updates. One useful acceleration technique is based on the 
parameter-expansion framework where one embeds the probability
model of interest within a larger probability model
that is identifiable from the complete data (\cite{Liu1998}). 
The resulting EM algorithm for the larger probability
model is referred to as a parameter-expanded EM (PX-EM) algorithm,
and in certain contexts, PX-EM algorithms can be developed
so that both the parameter updates of interest have a direct, closed 
form and the overall convergence speed is substantially improved.
Another variation of the EM algorithm which can aid more efficient computation
is the expectation-conditional maximization (ECM) algorithm (\cite{Meng1993}).
While not necessarily improving the rate of convergence, the ECM algorithm is a technique for simplifying the form of the parameter updates
by breaking the required maximization in the M-step
into a series of conditional maximization steps. 
A useful modification of ECM which can accelerate 
convergence is the expectation/conditional maximization either (ECME) algorithm (\cite{Liu1994}) which replaces one of the conditional maximization 
steps with a ``full" maximization step where one performs maximization with respect to the observed-data likelihood. For certain statistical models, parameter-expanded versions of ECME can be derived and these algorithms can be termed PX-ECME algorithms (\cite{Lewandowski2010}). PX-ECME algorithms have been used successfully in the context of parameter estimation for multivariate t-distributions (\cite{Liu1997}) and in fitting mixed effects models (\cite{vandyk2000}).
While not usually converging as quickly as a pure PX-EM algorithm, PX-ECME algorithms will, as argued in \cite{Lewandowski2010}, typically
converge faster than the EM algorithm, and hence, can be effective 
in cases when finding the full PX-EM parameter
update is cumbersome or infeasible.

The purpose of the current work is to explore the relative performance of a PX-ECME algorithm versus the associated EM algorithm for penalized weighted logistic regression with user-specified weights, and moreover, to show its connections with 
and relative performance compared to several other commonly
used optimization methods. 
We show that PX-ECME is indeed consistently faster than the associated EM algorithm
and that it can be somewhat competitive with the speed of Newton-Raphson in certain scenarios. In addition, we provide examples where Newton-Raphson often diverges whereas the PX-ECME algorithm provides quick, yet stable convergence to the maximizer of the weighted log-likelihood.

The organization of this article is as follows.
Section \ref{sec:em_description} reviews the P\'olya-Gamma representation
of a binomially distributed random variable with logistic link function
and describes the associated EM algorithm 
for weighted logistic regression. Next, 
Section \ref{sec:pxecme_description} reviews the construction of parameter-expanded EM algorithms and related PX-ECME algorithms. Then, Section \ref{sec:pxecme_description} outlines
a parameter-expanded version of the P\'olya-Gamma model and uses this parameter-expanded model to derive a PX-ECME algorithm for weighted logistic regression. Section \ref{sec:pxecme_description} also includes
results on the rates of convergence of the EM and PX-ECME algorithms.
Section \ref{sec:gpxem} outlines a more general 
PX-ECME algorithm and describes its connections to several other well-known procedures including proximal gradient descent \citep{Parikh2014} and
an MM algorithm for logistic regression \citep{bohning1988}.
Section \ref{sec:coord_descent} briefly describes a PX-ECME-type 
coordinate descent algorithm for logistic regression. 
Through a series of simulation studies, Section \ref{sec:sims} investigates the performance of our PX-ECME algorithms and compares its performance with several other procedures, and we then conclude with a brief
discussion in Section \ref{sec:conc}.

\section{Review of P\'olya-Gamma Data Augmentation for Logistic Regression} \label{sec:em_description}
Suppose we have $n$ responses $\mathbf{y} = (y_{1}, \ldots, y_{n})$ where
each $y_{i}$ is an integer satisfying $0 \leq y_{i} \leq m_{i}$, and for each $y_{i}$, we also have an associated covariate vector $\mathbf{x}_{i} \in \mathbb{R}^{p}$. A logistic regression model assumes that the $y_{i}$ are independent and that the distribution of each $y_{i}$ is given by 
\begin{equation}
y_{i} \sim \textrm{Binomial}\Big( m_{i}, \{ 1 + \exp(-\mathbf{x}_{i}^{T}\bbeta) \}^{-1} \Big), 
\label{eq:binomial_distribution}
\end{equation}
where $\bbeta = (\beta_{1}, \ldots, \beta_{p})^{T}$ is the vector of regression coefficients. It is common in practice to estimate $\bbeta$ by maximizing the observed-data log-likelihood $\ell_{o}(\bbeta|\mathbf{y})$ or a weighted observed log-likelihood function $\ell_{o, \mathbf{s}}(\bbeta|\mathbf{y})$. Under the assumed distribution (\ref{eq:binomial_distribution})
and a nonnegative vector of weights $\mathbf{s} = (s_{1}, \ldots, s_{n})$, the weighted observed log-likelihood function is given by
\begin{equation}
\ell_{o, \mathbf{s}}(\bbeta|\mathbf{y}) = \sum_{i=1}^{n}s_{i} \log \binom{m_{i}}{y_{i}} + \sum_{i=1}^{n} s_{i}y_{i}\mathbf{x}_{i}^{T}\bbeta - \sum_{i=1}^{n} s_{i}m_{i}\log\Big( 1 + \exp\{ \mathbf{x}_{i}^{T}\bbeta \} \Big).
\label{eq:observed_loglik}
\end{equation}

As outlined in \cite{Scott2013}, an EM algorithm for maximizing $\ell_{o, \mathbf{s}}(\bbeta|\mathbf{y})$ can be constructed by exploiting
the P\'olya-Gamma latent variable representation of the distribution of $y_{i}$ described in \cite{Polson2013}. A random variable $W$ is said to follow a P\'olya-Gamma distribution with parameters $b > 0$ and $c$ (denoted by $W \sim PG(b, c)$) if $W$ has the same distribution as an infinite convolution of independent Gamma-distributed random variables. Specifically, 
\begin{equation}
W  \stackrel{D}{=} \frac{1}{ 2\pi^{2} }\sum_{k=1}^{\infty} \frac{ G_{k} }{(k - 1/2)^{2} + c^{2}/(4\pi^{2})}, \nonumber 
\end{equation}
where $G_{k} \sim \textrm{Gamma}(b, 1)$ and $\stackrel{D}{=}$ denotes equality in distribution. The relevance of the P\'olya-Gamma family of distributions to logistic regression follows from the following key identity 
\begin{equation}
\frac{2^{m_{i}}\exp(m_{i}\mathbf{x}_{i}^{T}\bbeta/2)}{\{ 1 + \exp(\mathbf{x}_{i}^{T}\bbeta) \}^{m_{i}} } = \int_{0}^{\infty} \exp\big\{-w(\mathbf{x}_{i}^{T}\bbeta)^{2}/2 \big\} p_{PG}(w; m_{i}, 0) dw, \nonumber
\end{equation}
where $p_{PG}(w; b, c)$ denotes the density of a $PG(b, c)$ random variable and from the fact that the density of a $\textrm{PG}(m_{i}, \mathbf{x}_{i}^{T}\bbeta)$ random variable is given by
\begin{equation}
p_{PG}(w; m_{i}, \mathbf{x}_{i}^{T}\bbeta) = 2^{-m_{i}}\exp(-m_{i}\mathbf{x}_{i}^{T}\bbeta/2)\{1 + \exp(\mathbf{x}_{i}^{T}\bbeta) \}^{m_{i}}\exp\{ -w(\mathbf{x}_{i}^{T}\bbeta)^{2}/2 \}p_{PG}(w; m_{i}, 0). \nonumber
\end{equation}
Consequently, if one assumes that the observed responses $y_{1}, \ldots, y_{n}$ and latent random variables $W_{1}, \ldots, W_{n}$ arise from the following model 
\begin{eqnarray}
y_{i}|W_{i} &\sim& \textrm{Binomial}\Big( m_{i}, \{ 1 + \exp(-\mathbf{x}_{i}^{T}\bbeta)\}^{-1} \Big) \nonumber \\
W_{i} &\sim& PG(m_{i}, \mathbf{x}_{i}^{T}\bbeta), 
\label{eq:pg_representation}
\end{eqnarray}
then each $y_{i}$ has the correct marginal distribution, and the (weighted) complete-data log-likelihood $\ell_{c, \mathbf{s}}(\bbeta|\mathbf{v})$ corresponding to the complete data $\mathbf{v} = \{(y_{1}, W_{1}), \ldots, (y_{n}, W_{n})\}$ is
\begin{eqnarray}
&& \ell_{c, \mathbf{s}}(\bbeta|\mathbf{v}) = \nonumber \\
&& \sum_{i=1}^{n} s_{i}\log \binom{m_{i}}{y_{i}} + \sum_{i=1}^{n} s_{i} \log\Big( \frac{ \exp(y_{i}\mathbf{x}_{i}^{T}\bbeta)}{\{1 + \exp( \mathbf{x}_{i}^{T}\bbeta )\}^{m_{i}} } \Big) + \sum_{i=1}^{n} s_{i} \log\Big( \exp(-m_{i}\mathbf{x}_{i}^{T}\bbeta/2) \Big) \nonumber \\
&+& \sum_{i=1}^{n} s_{i} m_{i}\log\Big( 1 + \exp(\mathbf{x}_{i}^{T}\bbeta) \Big) - \frac{1}{2}\sum_{i=1}^{n} s_{i} W_{i}(\mathbf{x}_{i}^{T}\bbeta)^{2} + \sum_{i=1}^{n} s_{i} \log \{ p_{PG}(W_{i}; m_{i}, 0)/2^{m_{i}} \} \nonumber \\
&=& \sum_{i=1}^{n} s_{i}u_{i}\mathbf{x}_{i}^{T}\bbeta - \frac{1}{2}\sum_{i=1}^{n} s_{i} W_{i}(\mathbf{x}_{i}^{T}\bbeta)^{2} 
+ \sum_{i=1}^{n} s_{i} \Big( \log \binom{m_{i}}{y_{i} } + \log \big\{ p_{PG}(W_{i}; m_{i}, 0)/2^{m_{i}} \big\}\Big), 
\label{eq:complete_loglik}
\end{eqnarray}
where $u_{i} = y_{i} - m_{i}/2$.
In other words, the complete-data log-likelihood is a quadratic form in the vector of regression coefficients $\bbeta$. As noted in \cite{Scott2013}, this is very useful for constructing an EM algorithm as the M-step reduces to solving a weighted least-squares problem. The form of the complete-data log-likelihood is also useful in the context of Bayesian logistic regression (e.g., \cite{Polson2013}, \cite{choi2013}) as the conditional distribution of $\bbeta$ given the latent variables $W_{1}, \ldots, W_{n}$ is multivariate Gaussian.

To utilize the latent variable representation (\ref{eq:pg_representation}) to construct an EM algorithm for estimating $\bbeta$, one needs to find the expectation of $W_{i}$ given $y_{i}$ and current values of the regression coefficients.
Curiously, the distribution of $y_{i}$ does not depend on $W_{i}$ in (\ref{eq:pg_representation}), and hence, the expectations in the ``E-step" do not depend on the observed data. That is, $E\{ W_{i}| y_{i}, \bbeta^{(t)}\} = E\{ W_{i}|\bbeta^{(t)}\}$ for $i = 1, \ldots, n$, and 
as shown in \cite{Polson2013}, this expectation is given by 
\begin{equation}
\omega(\mathbf{x}_{i}^{T}\bbeta^{(t)}, m_{i}) = E\{ W_{i}|\bbeta^{(t)}\}
= \frac{m_{i}}{2\mathbf{x}_{i}^{T}\bbeta^{(t)}}\tanh\Big( \mathbf{x}_{i}^{T}\bbeta^{(t)}/2 \Big)
= \frac{m_{i}}{\mathbf{x}_{i}^{T}\bbeta^{(t)}}\Big( \frac{1}{1 + \exp(-\mathbf{x}_{i}^{T}\bbeta) } - \frac{1}{2} \Big), \nonumber
\end{equation}
with the understanding that, whenever $\mathbf{x}_{i}^{T}\bbeta^{(t)} = 0$, $\omega(\mathbf{x}_{i}^{T}\bbeta^{(t)}, m_{i})$ is set to $\lim_{y \longrightarrow 0} \tfrac{m_{i}}{2y}\tanh(\tfrac{y}{2}) = m_{i}/4$. It is useful to note the
following connection between the $\omega(\mathbf{x}_{i}^{T}\bbeta^{(t)}, m_{i})$ and the mean function $E(\mathbf{y}) = \mu(\mathbf{X}\bbeta)$
\begin{equation}
\mathbf{W}(\bbeta)\mathbf{X}\bbeta = \mu(\mathbf{X}\bbeta) - \tfrac{1}{2}\mathbf{m}, \quad \textrm{ where } \quad \mathbf{W}(\bbeta) = \textrm{diag}\big\{ \omega(\mathbf{x}_{1}^{T}\bbeta, m_{1}), \ldots, \omega( \mathbf{x}_{n}^{T}\bbeta, m_{n}) \big\}. 
\label{eq:meanfn_connection}
\end{equation}
In (\ref{eq:meanfn_connection}), $\mu(\mathbf{X}\bbeta) = E(\mathbf{y})$ is the $n \times 1$ vector whose $i^{th}$ component $[\mu(\mathbf{X}\bbeta)]_{i}$ is given by $[\mu(\mathbf{X}\bbeta)]_{i} = m_{i}/\{ 1 + \exp(\mathbf{x}_{i}^{T}\bbeta)\}$,
and $\mathbf{m}$ is the $n \times 1$ vector $\mathbf{m} = ( m_{1}, \ldots, m_{n} )^{T}$.

Parameter updates for the EM algorithm based on the latent variable model (\ref{eq:pg_representation}) are found by maximizing the ``Q-function" $Q(\bbeta|\bbeta^{(t)})$ which is defined as the expectation of $\ell_{c, \mathbf{s}}(\bbeta|\mathbf{x})$ given $\mathbf{y}$ and the current vector of parameter estimates $\bbeta^{(t)}$. From (\ref{eq:complete_loglik}), the Q-function is found by replacing $W_{i}$ with $\omega(\mathbf{x}_{i}^{T}\bbeta^{(t)}, m_{i})$ which allows the Q-function to be written as 
\begin{eqnarray}
Q(\bbeta|\bbeta^{(t)}) = E\{ \ell_{c, \mathbf{s}}(\bbeta|\mathbf{x})| \mathbf{y}, \bbeta^{(t)} \}
= -\frac{1}{2}\bbeta^{T}\mathbf{X}^{T}\mathbf{S}\mathbf{W}(\bbeta^{(t)})\mathbf{X}\bbeta + \bbeta^{T}\mathbf{X}^{T}\mathbf{S}\mathbf{u} + C, \nonumber
\end{eqnarray}
where $\mathbf{u} = (u_{1}, \ldots, u_{n})^{T}$, $\mathbf{W}(\bbeta^{(t)})$ is as defined in (\ref{eq:meanfn_connection}), $\mathbf{S} = \textrm{diag}\{ s_{1}, \ldots, s_{n} \}$, and $C$ is a constant not depending on $\bbeta$. The EM update $\bbeta^{(t+1)}$ of $\bbeta^{(t)}$ is found by maximizing $Q(\bbeta|\bbeta^{(t)})$ with respect to $\bbeta$, and hence we can represent $\bbeta^{(t+1)}$ as the solution of the following weighted least-squares problem 
\begin{eqnarray}
\bbeta^{(t+1)} &=& \argmin_{\boldsymbol{\beta} \in \mathbb{R}^{p}} 
\frac{1}{2}\Big(\mathbf{W}^{-1}(\bbeta^{(t)})\mathbf{u} - \mathbf{X}\bbeta \Big)^{T}\mathbf{S}\mathbf{W}(\bbeta^{(t)})\Big( \mathbf{W}^{-1}(\bbeta^{(t)})\mathbf{u} - \mathbf{X}\bbeta \Big) \nonumber \\
&=&(\mathbf{X}^{T}\mathbf{S}\mathbf{W}(\bbeta^{(t)})\mathbf{X})^{-1}\mathbf{X}^{T}\mathbf{S}\mathbf{u}. 
\label{eq:em_update}
\end{eqnarray}
It is interesting to note that the EM algorithm parameter update (\ref{eq:em_update}) is identical to the parameter update 
defined in \cite{jaakkola2000} which was based on variational Bayes arguments.
 
Based on (\ref{eq:em_update}), this EM algorithm for logistic regression can be viewed as an ``iteratively reweighted least squares" (IRLS) algorithm with weights $s_{1}\omega(\mathbf{x}_{1}^{T}\bbeta^{(t)}, m_{1}), \ldots, s_{n}\omega(\mathbf{x}_{n}^{T}\bbeta^{(t)}, m_{n})$ and ``response vector" $\mathbf{W}^{-1}(\bbeta^{(t)})\mathbf{u}$. In this sense, this EM algorithm closely resembles the common Newton-Raphson/Fisher scoring algorithm (\cite{Green1984}) used for logistic regression where one performs iteratively reweighted least squares using the weights $s_{i} \omega^{NR}(\mathbf{x}_{i}^{T}\bbeta^{(t)}, m_{i})$, where $\omega^{NR}(\mathbf{x}_{i}^{T}\bbeta^{(t)}, m_{i}) = m_{i}\exp(-\mathbf{x}_{i}^{T}\bbeta^{(t)})/[\{ 1 + \exp(-\mathbf{x}_{i}^{T}\bbeta^{(t)})\}^{2}]$. Note that, in the case of logistic regression, Newton-Raphson and Fisher scoring are equivalent methods, and we will refer to this procedure as Newton-Raphson in the remainder of the article.

\begin{figure}
\centering
     \includegraphics[width=6in,height=3.7in]{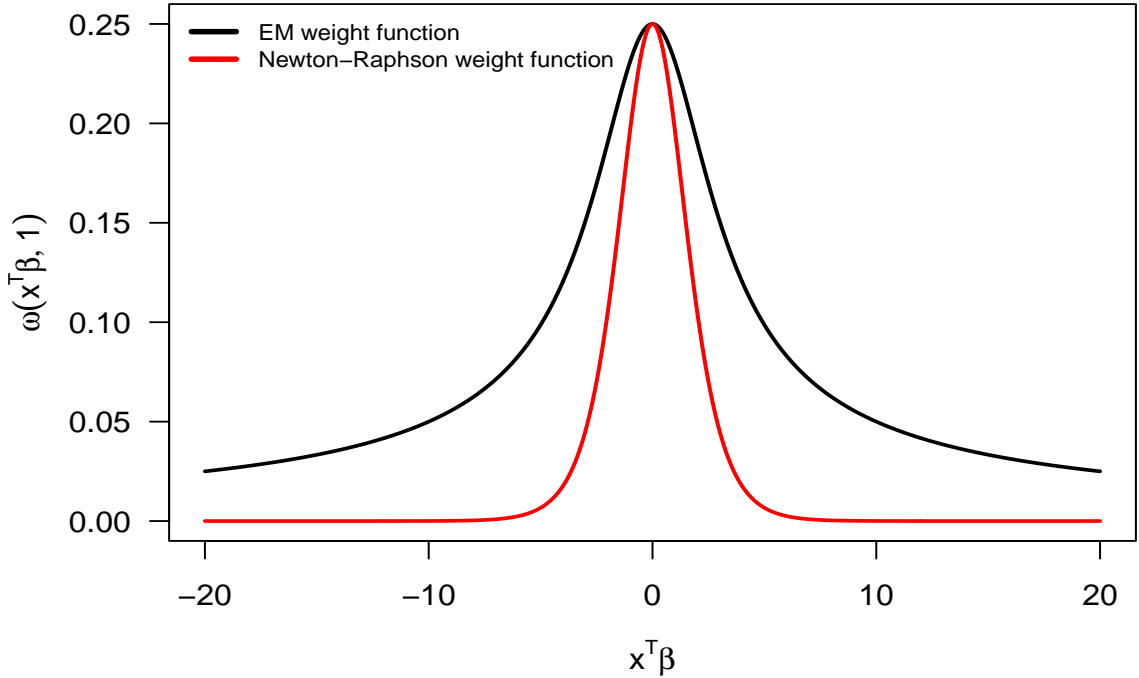}
\caption{(Color Figure) Weight functions $\omega(\mathbf{x}^{T}\bbeta, m)$ and $\omega^{NR}(\mathbf{x}^{T}\bbeta, m)$
used in the EM and Newton-Raphson algorithms respectively for $m=1$. This figure
shows the considerably heavier tails of the EM algorithm weight function.}
\label{fig:weight_functions}
\end{figure}

To better see the resemblance between the EM and Newton-Raphson updates, it interesting to note that we can also express the EM update (\ref{eq:em_update}) of $\bbeta^{(t)}$ as
\begin{equation}
\bbeta^{(t+1)} 
= \bbeta^{(t)} + (\mathbf{X}^{T}\mathbf{S}\mathbf{W}(\bbeta^{(t)})\mathbf{X})^{-1}\mathbf{X}^{T}\mathbf{S}\{\mathbf{y} - \mu(\mathbf{X}\bbeta^{(t)})\}.  \nonumber
\end{equation}
The above expression for the EM update is a consequence of the equality $\mathbf{W}(\bbeta)\mathbf{X}\bbeta = \mu(\mathbf{X}\bbeta) - \tfrac{1}{2}\mathbf{m}$ stated in (\ref{eq:meanfn_connection}).
The main difference between the EM weight function $\omega(\mathbf{x}^{T}\bbeta, m)$ and the Newton-Raphson weight function $\omega^{NR}(\mathbf{x}^{T}\bbeta, m)$ is that, for a fixed value of $m$, $\omega(\mathbf{x}^{T}\bbeta, m)$ has much heavier tails than $\omega^{NR}(\mathbf{x}^{T}\bbeta, m)$. This is illustrated in Figure \ref{fig:weight_functions} which plots $\omega(\mathbf{x}^{T}\bbeta, m)$ and $\omega^{NR}(\mathbf{x}^{T}\bbeta, m)$ when it is assumed that $m = 1$.

A main advantage of the EM algorithm over the Newton-Raphson algorithm is that the EM algorithm is monotone; namely, iterates from the EM algorithm are guaranteed to increase the weighted log-likelihood at every iteration and the iterates are guaranteed to converge to a fixed-point of the algorithm. It is interesting to note that the first iterations of the EM and Newton-Raphson are identical whenever all components of the initial iterate $\bbeta^{(0)}$ are set to zero. This is a consequence of the fact that $\omega^{EM}(0, m_{i}) = \omega^{NR}(0, m_{i}) = m_{i}/4$. While the Newton-Raphson algorithm is not monotone and does not possess any global convergence guarantees, the fact that the first step of Newton-Raphson increases the value of the log-likelihood when setting $\bbeta^{(0)} = \mathbf{0}$ is one reason for the relatively robust performance of Newton-Raphson for logistic regression. The monotonicity of the Newton-Raphson first step when setting $\bbeta^{(0)} = \mathbf{0}$ was also noted in \cite{bohning1988}.

\section{A Parameter-Expanded ECME Algorithm for Logistic Regression} \label{sec:pxecme_description}
\subsection{Review of Parameter-Expanded EM and ECME Algorithms} \label{ss:px_review}
In a variety of estimation problems, the convergence speed of the EM algorithm can be improved by viewing the complete-data model
as embedded within a larger model that has additional parameters and then performing parameter updates with respect to the larger model. An EM algorithm constructed from a larger, parameter-expanded model is typically referred to as a parameter-expanded EM algorithm (PX-EM) algorithm (\cite{Liu1998}).

To describe the steps involved in the PX-EM algorithm, we consider a general setup where one is interested in estimating a parameter vector $\bbeta \in \mathcal{B}$ and one has defined a complete-data vector $\mathbf{v}$ with density function $g_{c}(\mathbf{v}|\bbeta)$. The complete-data vector $\mathbf{v} \in \mathcal{V}$ and observed-data vector $\mathbf{y} \in \mathcal{Y}$ are connected through $\mathbf{y} = h(\mathbf{v})$ where $h: \mathcal{V} \longrightarrow \mathcal{Y}$ is the data-reducing mapping to be used for developing the EM algorithm.
As described in \cite{Liu1998}, to generate a PX-EM algorithm one must first consider an expanded probability model for the complete-data vector $\mathbf{v}$, and we let $g_{PX}(\mathbf{v}|\btheta, \alpha)$ denote the probability density function
for this expanded probability model where $(\btheta, \alpha)$ are the parameters of the expanded model and $\Theta$ denotes the parameter space for $(\btheta, \alpha)$. As outlined in \cite{Liu1998} and \cite{Lewandowski2010},
the expanded probability model should satisfy the following conditions in order to generate a well-defined PX-EM algorithm
\begin{enumerate}
    \item 
    The observed-data model is preserved in the sense that there is a many-to-one reduction function $R: \Theta \longrightarrow \mathcal{B}$ such that 
    \begin{equation}
        g_{O}\big\{\mathbf{y}|\bbeta = R(\btheta, \alpha) \big\} = \int_{\mathcal{V}(\mathbf{y})} g_{PX}(\mathbf{v}|\btheta, \alpha) d\mathbf{v}, \quad \textrm{ for all } (\btheta, \alpha) \in \Theta, \nonumber
    \end{equation}
    where $\mathcal{V}(\mathbf{y}) = \{ \mathbf{v} \in \mathcal{V}: h(\mathbf{v}) = \mathbf{y} \}$.
    \item
    There is a ``null value" $\alpha_{0}$ of $\alpha$ such that the complete-data model is preserved at the null value in the sense that
    \begin{equation}
        g_{PX}(\mathbf{v}|\bbeta, \alpha_{0}) = g_{c}(\mathbf{v}|\bbeta), \qquad \textrm{for all } \bbeta \in \mathcal{B}. \nonumber
    \end{equation}
\end{enumerate}
Similar to the EM algorithm, the PX-EM algorithm consists of two steps: a ``PX--E" step followed by a ``PX--M" step. 
The PX--E step is similar to the usual ``E--step" of an EM algorithm in that a ``Q-function" is formed by taking the expectation of the (parameter-expanded) complete-data log likelihood conditional on the observed data and parameter values $(\btheta, \alpha) = (\bbeta^{(t)}, \alpha_{0})$. The ``PX--M" step then proceeds by finding the expanded parameter values $(\btheta^{(t+1)}, \alpha^{(t+1)})$ which maximize this Q-function and computing the parameter update $\bbeta^{(t+1)}$ by applying the reduction function to $(\btheta^{(t+1)}, \alpha^{(t+1)})$. The two steps of the PX-EM algorithm can be summarized as:
\begin{enumerate} \itemsep4pt
    \item 
    \textit{PX--E step:} Compute the parameter-expanded Q-function
    \begin{equation}
        Q_{PX}(\btheta, \alpha| \bbeta^{(t)}, \alpha_{0}) = E\Big\{ \log g_{PX}(\mathbf{v}|\btheta, \alpha) \Big| \mathbf{y}, \btheta = \bbeta^{(t)}, \alpha = \alpha_{0} \Big\}. 
        \label{eq:px_Qfunction}
    \end{equation}
    \item
    \textit{PX--M step:} Find
    \begin{equation}
        (\btheta^{(t+1)}, \alpha^{(t+1)}) = \argmax_{\boldsymbol{\theta, \alpha}} Q_{PX}(\btheta, \alpha| \bbeta^{(t)}, \alpha_{0}), \nonumber 
    \end{equation}
    and update $\bbeta^{(t)}$ using
    \begin{equation}
        \bbeta^{(t+1)} = R(\btheta^{(t+1)}, \alpha^{(t+1)}). \nonumber 
    \end{equation}
\end{enumerate}

A chief advantage of the PX-EM algorithm is that it is guaranteed to converge at least as fast
as the corresponding EM algorithm. When the parameter-expanded model is constructed so that
all parameter updates have a direct, closed-form, the PX-EM algorithm can have substantially faster
convergence speed than the EM algorithm while maintaining the stability
and computational simplicity of the original EM algorithm.

The Expectation-Conditional Maximization (ECM) algorithm (\cite{Meng1993}) and the Expectation-Conditional Maximization Either (ECME) algorithms (\cite{Liu1994}) are variations of the EM algorithm where the M-step is replaced by a sequence of conditional maximization steps. In the ECM algorithm, one breaks up the maximization of the Q-function into several conditional maximization steps where, in each conditional maximization step, one only maximizes the Q-function with respect to a subset of the parameters while keeping the remaining parameter values fixed. In the ECME algorithm, one also performs a series of conditional maximization steps. However, in the ECME algorithm, some of the conditional maximization are performed with respect to the observed-data log-likelihood function while the remaining conditional maximization steps are performed with respect to the usual Q-function.

As in the original ECME algorithm, in a parameter-expanded version of ECME (PX-ECME), one performs a sequence of conditional maximization steps for the expanded parameters $(\btheta, \alpha)$ with some of the
conditional maximization steps being performed with respect to the parameter-expanded Q-function 
while the other conditional maximization steps are performed with respect to the observed-data log-likelihood function.
Specifically, when using the observed-data log-likelihood function for conditional maximization one will maximize $\ell_{o, \mathbf{s}}\{ R(\btheta, \alpha)|\mathbf{y} \}$ with respect to the subset of parameters being maximized in
that conditional maximization step. After all of the expanded parameters have been updated through
the sequence of conditional maximization steps, the update for the parameter of interest $\bbeta$
is found by applying the reduction function $R(\cdot)$ to the updated expanded parameters.

One example of a PX-ECME algorithm is the procedure described in \cite{Liu1997} for estimating the parameters
of a multivariate-t distribution. Here, the EM algorithm exploits the representation of the multivariate-t distribution
as a scale mixture of multivariate normal distributions, and the parameter-expanded model utilized by \cite{Liu1997}
includes an expanded scale parameter that plays a role in both the covariance matrix of the normal distribution
and the scale of the latent Gamma distribution. This expanded parameter is updated in each 
iteration through a conditional maximization step maximizing the observed-data likelihood.

One point worth noting is that a PX-ECME algorithm can often be constructed even if the expanded parameter vector $(\btheta, \alpha)$ is not identifiable from the complete data. A PX-ECME algorithm can work
if the subsets of parameters used in each conditional maximization step are conditionally identifiable from the observed data. Specifically, the parameters in each of the subsets
should be identifiable from the observed data assuming that all the other parameters are fixed.
For example, if we construct our PX-ECME algorithm so that $\btheta$ is updated first followed by an update of $\alpha$,
then $\btheta$ should be identifiable for a fixed value of $\alpha$, and $\alpha$ should be identifiable from the observed data for a fixed value of $\btheta$.

\subsection{A Parameter-Expanded ECME Algorithm for Logistic and Penalized Logistic Regression} \label{ss:pxecme_logistic}
We consider the following parameter-expanded version of model (\ref{eq:pg_representation}) for the complete-data vector $\mathbf{v} = \{(y_{1}, W_{1}), \ldots, (y_{n}, W_{n})\}$
\begin{eqnarray}
y_{i}|W_{i} &\sim& \textrm{Binomial}\Big( m_{i}, \{ 1 + \exp(-\alpha\mathbf{x}_{i}^{T}\btheta)\}^{-1} \Big) \nonumber \\
W_{i} &\sim& PG(m_{i}, \alpha\mathbf{x}_{i}^{T}\btheta), 
\label{eq:pgexpanded_representation}
\end{eqnarray}
where the parameters in this model are
$\btheta \in \mathbb{R}^{p}$ and $\alpha \in \mathbb{R}$. Under this expanded model, the observed-data model is preserved when using the reduction function $\bbeta = R(\btheta, \alpha) = \alpha\btheta$, and the original complete-data model (\ref{eq:pg_representation}) is preserved whenever $\alpha = \alpha_{0} = 1$.

We now turn to describing how to use the parameter-expanded model (\ref{eq:pgexpanded_representation}) to construct a PX-ECME algorithm for maximizing a \textit{penalized} weighted log-likelihood function with penalty function $P_{\boldsymbol{\eta}}(\bbeta)$. 
A PX-ECME algorithm for maximizing a penalized log-likelihood is very similar to that of maximizing a log-likelihood
function the only difference being that a penalty function will be subtracted from either the parameter-expanded Q-function
or the observed-data log-likelihood function.
Throughout the remainder of this paper, we will assume the penalty function has the form $P_{\boldsymbol{\eta}}(\bbeta) = \sum_{j=1}^{p} P_{\boldsymbol{\eta}, j}(\beta_{j})$, where $\boldeta \in \mathbb{R}^{q}$ is a vector of tuning parameters. 
The observed-data weighted penalized log-likelihood function $p\ell_{o, \mathbf{s}}^{\boldsymbol{\eta}}$  that we seek to maximize is then
\begin{equation}
p\ell_{o, \mathbf{s}}^{\boldsymbol{\eta}}(\bbeta|\mathbf{y}) = \sum_{i=1}^{n} s_{i}\log \binom{m_{i}}{y_{i}} + \sum_{i=1}^{n} s_{i} y_{i} \mathbf{x}_{i}^{T}\bbeta - \sum_{i=1}^{n} s_{i}m_{i}\log\Big( 1 + \exp\{ \mathbf{x}_{i}^{T}\bbeta \} \Big) - P_{\boldsymbol{\eta}}(\bbeta), \nonumber
\end{equation}
and the complete-data penalized weighted log-likelihood corresponding to model (\ref{eq:pgexpanded_representation}) is
\begin{equation}
p\ell_{c,PX}^{\boldsymbol{\eta}}(\btheta, \alpha| \mathbf{v})
= \alpha\sum_{i=1}^{n} s_{i} u_{i}\mathbf{x}_{i}^{T}\btheta - \frac{\alpha^{2}}{2}\sum_{i=1}^{n} s_{i} W_{i}(\mathbf{x}_{i}^{T}\btheta)^{2} - P_{\boldsymbol{\eta}}(\alpha\btheta) + C, \nonumber
\end{equation}
where $C$ is a constant that does not depend on $(\btheta, \alpha)$.
The penalized, parameter-expanded Q-function $Q_{PX}^{\boldsymbol{\eta}}(\btheta, \alpha|\bbeta^{(t)}, \alpha_{0})$ is obtained by just subtracting $P_{\boldsymbol{\eta}}(\alpha \btheta)$ from the usual parameter-expanded Q-function (\ref{eq:px_Qfunction}). For model (\ref{eq:pgexpanded_representation}), this is given by 
\begin{equation}
Q_{PX}^{\boldsymbol{\eta}}(\btheta, \alpha|\bbeta^{(t)}, \alpha_{0}) 
= \alpha \btheta^{T}\mathbf{X}^{T}\mathbf{S}\mathbf{u} - \frac{\alpha^{2}}{2}\btheta^{T}\mathbf{X}^{T}\mathbf{S}\mathbf{W}(\bbeta^{(t)})\mathbf{X}\btheta - P_{\boldsymbol{\eta}}(\alpha\btheta) + C,
\label{eq:px_qfunction}
\end{equation}
where $\mathbf{W}(\bbeta^{(t)})$ is the $n \times n$ weight matrix defined in (\ref{eq:meanfn_connection}).

One feature of the parameter-expanded model (\ref{eq:pgexpanded_representation}) is that the parameter vector $(\btheta, \alpha)$ is not identifiable from the complete-data vector $\mathbf{v}$. This may be seen by noting that, for a positive scalar c, the parameters $(\alpha, \btheta)$ and $(\alpha/c, c\btheta)$ both lead to the same complete-data penalized log-likelihood. Nevertheless, $(\btheta, \alpha)$ are conditionally identifiable meaning that, for a fixed $\btheta$, $\alpha$ is identifiable, and for a fixed $\alpha$, $\btheta$ is identifiable. Hence, we can first update $\btheta^{(t)}$ by fixing $\alpha = \alpha^{(t)}$ and maximizing $Q_{PX}^{\boldsymbol{\eta}}(\btheta, \alpha|\bbeta^{(t)}, \alpha_{0})$ with respect to $\btheta$
\begin{eqnarray}
\btheta^{(t+1)} &=& \argmax_{\boldsymbol{\theta} \in \mathbb{R}^{p}} Q_{PX}^{\boldsymbol{\eta}}(\btheta, \alpha^{(t)}|\bbeta^{(t)}, \alpha_{0}) \nonumber \\
&=& \argmin_{\boldsymbol{\theta} \in \mathbb{R}^{p}} \Bigg[ -\alpha^{(t)} \btheta^{T}\mathbf{X}^{T}\mathbf{S}\mathbf{u} + \frac{(\alpha^{(t)})^{2}}{2}\btheta^{T}\mathbf{X}^{T}\mathbf{S}\mathbf{W}(\bbeta^{(t)})\mathbf{X}\btheta + P_{\boldsymbol{\eta}}(\alpha^{(t)}\btheta) - C \Bigg]. 
\label{eq:px_update} 
\end{eqnarray}
We then update $\alpha^{(t)}$ by maximizing the observed-data penalized log-likelihood 
\begin{equation}
\alpha^{(t+1)} = \argmax_{\alpha \in \mathbb{R}} p\ell_{o}^{\boldsymbol{\eta}}(\alpha\btheta^{(t+1)}|\mathbf{y}). \nonumber 
\end{equation}
The PX-ECME parameter update of $\bbeta^{(t)}$ is given by $\bbeta^{(t+1)} = R(\btheta^{(t+1)}, \alpha^{(t+1)}) = \alpha^{(t+1)}\btheta^{(t+1)}$.

It is worth noting that one can express $\btheta^{(t+1)}$ in terms of the EM parameter update as $\btheta^{(t+1)} = \bbeta^{(t+1),  EM}/\alpha^{(t)}$ where the EM update $\bbeta^{(t+1), EM}$ is 
obtained from $\bbeta^{(t)}$ by maximizing $Q_{PX}^{\boldsymbol{\eta}}(\btheta, 1|\bbeta^{(t)}, 1)$ with 
respect to $\btheta$. Hence, we can express the PX-ECME update as
\begin{equation}
\bbeta^{(t+1)} = \frac{1}{\alpha^{(t)}} \bbeta^{(t+1),EM} \Big[ \argmax_{\alpha \in \mathbb{R}} p\ell_{o, \mathbf{s}}^{\boldsymbol{\eta}}(\alpha\bbeta^{(t+1), EM}/\alpha^{(t)}|\mathbf{y}) \Big]
= \rho^{(t+1)}\bbeta^{(t+1),EM}, \nonumber
\end{equation}
where $\rho^{(t+1)} = \argmax_{\rho \in \mathbb{R} } p\ell_{o,\mathbf{s}}^{\boldsymbol{\eta}}( \rho \bbeta^{(t+1), EM}|\mathbf{y})$. In other 
words, the PX-ECME update is found by simply first computing the
EM update $\bbeta^{(t+1), EM}$ of the regression coefficients and then multiplying $\bbeta^{(t+1), EM}$
by the scaling factor $\rho^{(t+1)}$, where $\rho^{(t+1)}$ is the scalar maximizing the weighted observed-data penalized
log-likelihood function among regression coefficients of the form $\rho\bbeta^{(t+1), EM}$. The 
complete PX-ECME procedure is summarized in Algorithm \ref{alg:basic_pxecme}. 

Finding $\rho^{(t+1)}$ only involves solving a one-dimensional root-finding problem, namely, 
$\partial p\ell_{o,\mathbf{s}}^{\boldsymbol{\eta}}(\rho \bbeta^{(t+1), EM}|\mathbf{y})/\partial \rho = 0$, of which there 
are a wide range of available numerical methods, and in such cases,
robust root-finding methods such as Brent's method (\cite{brent2013}) may be directly used to find $\rho^{(t+1)}$.

\begin{algorithm}[H]
\setstretch{1.25}
\caption{(PX-ECME for Penalized Logistic Regression). 
In the description of the algorithm, $\mathbf{S} = \textrm{diag}\{ s_{1}, \ldots, s_{n}\}$ and
$\mathbf{u} = (u_{1}, \ldots, u_{n})^{T}$, where $u_{i} = y_{i} - m_{i}/2$.}
\begin{algorithmic}[1]
\State Given $\bbeta^{(0)} \in \mathbb{R}^{p}$.
\For{t=0,1,2,... until convergence}
\State Compute the $n \times n$ diagonal weight matrix $\mathbf{W}(\bbeta^{(t)})$ whose
$i^{th}$ diagonal element is
\begin{equation}
\omega(\mathbf{x}_{i}^{T}\bbeta, m_{i}) = \frac{m_{i}}{2\mathbf{x}_{i}^{T}\bbeta^{(t)}}\tanh\big( \mathbf{x}_{i}^{T}\bbeta^{(t)}/2 \big). \nonumber
\end{equation}
\State Compute $\bbeta^{(t+1), EM}$:
\vspace{-.2cm}
\begin{equation}
\bbeta^{(t+1), EM} = \argmin_{\boldsymbol{\beta} \in \mathbb{R}^{p}} \Big[ \frac{1}{2}\bbeta^{T}\mathbf{X}^{T}\mathbf{S}\mathbf{W}(\bbeta^{(t)})\mathbf{X}\bbeta - \bbeta^{T}\mathbf{X}^{T}\mathbf{S}\mathbf{u} + P_{\boldsymbol{\eta}}( \bbeta ) \Big]. \nonumber
\vspace{-.2cm}
\end{equation}
\State Compute
\begin{equation}
\rho^{(t+1)} = \argmax_{\rho \in \mathbb{R} } \Big[ p\ell_{o,\mathbf{s}}^{\boldsymbol{\eta}}( \rho \bbeta^{(t+1), EM}|\mathbf{y}) \Big].  \nonumber
\end{equation}
\State Set $\bbeta^{(t+1)} = \rho^{(t+1)}\bbeta^{(t+1), EM}$.
\EndFor
\end{algorithmic}
\label{alg:basic_pxecme}
\end{algorithm}

An attractive feature of the PX-ECME algorithm (Algorithm \ref{alg:basic_pxecme}) is that it is, like the EM algorithm, a monotone algorithm. The monotonicity property helps to ensure that the procedure stable and robust in practice. The monotonicity of PX-ECME directly follows from the fact that the EM algorithm is monotone with respect to the penalized log-likelihood
(i.e., $p\ell_{o,\mathbf{s}}^{\boldsymbol{\eta}}(\bbeta^{(t+1),EM}|\mathbf{y}) \geq p\ell_{o,\mathbf{s}}^{\boldsymbol{\eta}}(\bbeta^{(t)}|\mathbf{y})$) and that $p\ell_{o,\mathbf{s}}^{\boldsymbol{\eta}}(\rho^{(t+1)}\bbeta^{(t+1),EM}|\mathbf{y}) \geq p\ell_{o,\mathbf{s}}^{\boldsymbol{\eta}}(\bbeta^{(t+1), EM}|\mathbf{y})$.


\subsection{Monotone acceleration of EM via order-1 Anderson acceleration} \label{ss:aa_overview}
The PX-ECME algorithm accelerates convergence by multiplying the EM parameter update by a single scalar, and there are likely other
straightforward modifications of the EM algorithm that
can produce substantial speed-ups in convergence without sacrificing
the monotonicity property. 

An attractive class of methods with straightforward parameter updates are those that simply take a linear combination 
of the previous two EM parameter updates with the weights in
the linear combination determined adaptively. One can ensure that this scheme is monotone by comparing the value of the objective function for a parameter update versus that of the current parameter value. One method
for finding the weights in this linear combination is the order-1 Anderson acceleration (\cite{Walker2011}). 
Applying the order-1 Anderson acceleration to the EM algorithm for penalized logistic regression leads to the following parameter updating scheme
\begin{equation}
\bbeta^{(t+1)} = 
\begin{cases} 
(1 - \gamma^{(t)}) \bbeta^{(t+1), EM} + \gamma^{(t)} \bbeta^{(t), EM} & \text{ if $p\ell_{o, \mathbf{s}}^{\boldsymbol{\eta}}(\bbeta^{(t+1)}|\mathbf{y}) \geq p\ell_{o, \mathbf{s}}^{\boldsymbol{\eta}}(\bbeta^{(t+1),EM}|\mathbf{y})$  } \nonumber \\
\bbeta^{(t+1),EM} & \text{ otherwise, }
\end{cases}
\end{equation}
where $\gamma^{(t)}$ is the scalar $\gamma^{(t)} = \mathbf{v}_{t}^{T}\mathbf{r}_{t}/\mathbf{v}_{t}^{T}\mathbf{v}_{t}$ and where
$\mathbf{r}_{t} = \bbeta^{(t+1), EM} - \bbeta^{(t)}$ and $\mathbf{v}_{t} = \bbeta^{(t+1), EM} - \bbeta^{(t)} + \bbeta^{(t-1)} - \bbeta^{(t),EM}$.
Because the underlying EM algorithm is monotone, only accepting the linear combination $\bbeta^{(t+1)}$ if $p\ell_{o, \mathbf{s}}^{\boldsymbol{\eta}}(\bbeta^{(t+1)}|\mathbf{y}) \geq p\ell_{o, \mathbf{s}}^{\boldsymbol{\eta}}(\bbeta^{(t+1),EM}|\mathbf{y})$
ensures that the above iterative scheme is monotone. 

\subsection{Rate of Convergence of EM and PX-ECME for Logistic Regression} \label{ss:convergencerate}

The PX-ECME updating scheme outlined in Algorithm \ref{alg:basic_pxecme} implicitly defines a mapping $G: \mathbb{R}^{p} \longrightarrow \mathbb{R}^{p}$
which performs the PX-ECME update of $\bbeta^{(t)}$, i.e., $\bbeta^{(t+1)} = G_{PX}(\bbeta^{(t)})$. Likewise, the EM algorithm
also defines a mapping $\bbeta^{(t+1), EM} = G_{EM}(\bbeta^{(t), EM})$ which, from (\ref{eq:em_update}), has the form $G_{EM}(\bbeta) = (\mathbf{X}^{T}\mathbf{S}\mathbf{W}(\bbeta)\mathbf{X})^{-1}\mathbf{X}^{T}\mathbf{S}\mathbf{u}$.
Because the PX-ECME parameter update is a scalar multiple of the EM update, we can express the 
mapping $G_{PX}$ in terms of $G_{EM}$ as
\begin{equation}
G_{PX}(\bbeta) = h\{ G_{EM}(\bbeta) \}G_{EM}(\bbeta),
\end{equation}
where $h: \mathbb{R}^{p} \longrightarrow \mathbb{R}$ is the function defined as
$h(\bbeta) = \argmax_{\rho} [p\ell_{o, \mathbf{s}}^{\boldsymbol{\eta}}(\rho\bbeta|\mathbf{y})]$.

The rate of convergence of a fixed-point iteration of the form $\bbeta^{(t+1)} = G(\bbeta^{(t)})$ is typically
defined as the maximum eigenvalue of the Jacobian matrix of $G$ when the Jacobian is evaluated at the 
convergence point $\bbeta^{*}$ of the algorithm (see, e.g., \cite{mclachlan2007}).
As shown in \cite{durante2018}, the Jacobian matrix $\mathbf{J}_{EM}(\bbeta)$ associated with the mapping 
$G_{EM}(\bbeta) = (\mathbf{X}^{T}\mathbf{S}\mathbf{W}(\bbeta)\mathbf{X})^{-1}\mathbf{X}^{T}\mathbf{S}\mathbf{u}$ is given by
\begin{equation}
\mathbf{J}_{EM}(\bbeta) = \mathbf{I}_{p} - (\mathbf{X}^{T}\mathbf{S}\mathbf{W}(\bbeta)\mathbf{X})^{-1}\mathbf{X}^{T}\mathbf{S}\mathbf{R}(\bbeta)\mathbf{X},
\end{equation}
where $\mathbf{R}(\bbeta)$ is the $n \times n$ diagonal matrix whose $i^{th}$ diagonal element 
is $m_{i}\pi(\mathbf{x}_{i}^{T}\bbeta)[1 - \pi(\mathbf{x}_{i}^{T}\bbeta)]$
and $\pi(u) = 1/\{1 + \exp(-u)\}$.
As we show in Theorem \ref{thm:rate_of_convergence} below, the 
maximum eigenvalue of the Jacobian matrix of $G_{PX}$ at $\bbeta^{*}$ can be
no larger than the maximum eigenvalue of $\mathbf{J}_{EM}(\bbeta^{*})$, and hence the rate of convergence of PX-ECME is no slower than the EM algorithm.

\begin{theorem}
When assuming the penalty function equals zero for all $\bbeta$ (i.e., $P_{\boldsymbol{\eta}}(\bbeta) \equiv 0$) and $\bbeta^{*} \neq \mathbf{0}$, the Jacobian $\mathbf{J}_{PX}(\bbeta) = \partial G_{PX}(\bbeta)/\partial \bbeta$ of the PX-ECME mapping at $\bbeta = \bbeta^{*}$ is given by
\begin{eqnarray}
\mathbf{J}_{PX}(\bbeta^{*}) &=& 
\mathbf{I}_{p} - \mathbf{E}^{-1}(\bbeta^{*})\mathbf{V}(\bbeta^{*}) - [c(\bbeta^{*})]^{-1}\bbeta^{*}(\bbeta^{*})^{T}\mathbf{V}(\bbeta^{*})\big\{ \mathbf{V}^{-1}(\bbeta^{*}) - \mathbf{E}^{-1}(\bbeta^{*}) \big\}\mathbf{V}(\bbeta^{*})^{T} \nonumber \\
&=& \mathbf{J}_{EM}(\bbeta^{*}) - [c(\bbeta^{*})]^{-1}\bbeta^{*}(\bbeta^{*})^{T}\mathbf{V}(\bbeta^{*})\big\{ \mathbf{V}^{-1}(\bbeta^{*}) - \mathbf{E}^{-1}(\bbeta^{*}) \big\}\mathbf{V}(\bbeta^{*})^{T}, \nonumber
\end{eqnarray}
where $\mathbf{J}_{EM}(\bbeta^{*})$ is the Jacobian matrix of the EM mapping at $\bbeta = \bbeta^{*}$, 
$c(\bbeta^{*}) = (\bbeta^{*})^{T}\mathbf{X}^{T}\mathbf{S}\mathbf{R}(\bbeta^{*})\mathbf{X}\bbeta^{*}$,
$\mathbf{E}(\bbeta^{*}) = \mathbf{X}^{T}\mathbf{S}\mathbf{W}(\bbeta^{*})\mathbf{X}$, and $\mathbf{V}(\bbeta^{*}) = \mathbf{X}^{T}\mathbf{S}\mathbf{R}(\bbeta^{*})\mathbf{X}$. 
In addition, we have $r_{PX} \leq r_{EM}$, where $r_{PX}$ and $r_{EM}$ are the maximal 
eigenvalues of $\mathbf{J}_{PX}(\bbeta^{*})$ and $\mathbf{J}_{EM}(\bbeta^{*})$ respectively.
\label{thm:rate_of_convergence}
\end{theorem}

\section{A Generalized PX-ECME Algorithm for Logistic Regression} \label{sec:gpxem}

\subsection{Description of the Aproach}
For many penalty functions, the maximization in (\ref{eq:px_update}) does not yield an easily-computable solution and may require the use of a time-consuming iterative procedure to compute $\btheta^{(t+1)}$. To better handle such cases, we consider a generalized PX-ECME algorithm where $\btheta^{(t+1)}$ maximizes a more manageable, surrogate parameter-expanded Q-function.
The more manageable surrogate Q-functions $\tilde{Q}_{PX}^{\boldsymbol{\eta}}(\btheta, \alpha|\bbeta^{(t)}, \alpha_{0})$ are assumed to have the form
\begin{eqnarray}
&& \tilde{Q}_{PX}^{\boldsymbol{\eta}}(\btheta, \alpha|\bbeta^{(t)}, \alpha_{0})
 =  Q_{PX}^{\boldsymbol{\eta}}(\btheta, \alpha|\bbeta^{(t)}, \alpha_{0}) - \frac{1}{2}(\alpha\btheta - \bbeta^{(t)})^{T}\mathbf{H}^{(t)}(\alpha\btheta - \bbeta^{(t)}) \nonumber \\
&& = \alpha\btheta^{T}[\mathbf{X}^{T}\mathbf{S}\mathbf{u} + \mathbf{H}^{(t)}\bbeta^{(t)}] - \frac{\alpha^{2}}{2}\btheta^{T}\big[ \mathbf{X}^{T}\mathbf{S}\mathbf{W}(\bbeta^{(t)})\mathbf{X} + \mathbf{H}^{(t)} \big]\btheta - P_{\boldsymbol{\eta}}(\alpha\btheta) + C, 
\label{eq:surrogateQ_form}
\end{eqnarray}
where $\mathbf{H}^{(t)}$ is a positive semi-definite $p \times p$ matrix. As in the PX-ECME algorithm for penalized logistic regression, to update $\bbeta^{(t)}$ one first finds $\btheta^{(t+1)}$ and $\alpha^{(t+1)}$ by performing two conditional maximization steps with respect to $\tilde{Q}_{PX}^{\boldsymbol{\eta}}$ and then sets $\bbeta^{(t+1)} = \alpha^{(t+1)}\btheta^{(t+1)}$. We will refer to an algorithm which updates $\bbeta^{(t)}$ in this manner as a \textit{generalized parameter-expanded} ECME (GPX-ECME) algorithm for penalized weighted logistic regression.

The steps involved in a GPX-ECME are outlined in Algorithm \ref{alg:gpx_ecme}. Notice in this Algorithm that, as in the case of a PX-ECME algorithm,
the process of first computing $\btheta^{(t+1)}$ and $\alpha^{(t+1)}$ and then setting $\bbeta^{(t+1)} = \alpha^{(t+1)}\btheta^{(t+1)}$ may be equivalently expressed as $\bbeta^{(t+1)} = \rho^{(t+1)}\bbeta^{(t+1),GEM}$, where $\bbeta^{(t+1), GEM}$ maximizes $\tilde{Q}_{PX}^{\boldsymbol{\eta}}(\bbeta, 1|\bbeta^{(t)}, \alpha_{0})$ and $\rho^{(t+1)}$ maximizes $p\ell_{o, \mathbf{s}}^{\boldsymbol{\eta}}(\rho \bbeta^{(t+1), GEM} | \mathbf{y} )$ with respect to $\rho \in \mathbb{R}$.
As stated in the following proposition, as long as $\mathbf{H}^{(t)}$ is a positive semi-definite matrix for each $t$, each iterate generated from a GPX-ECME
algorithm will increase the value of the penalized log-likelihood. 

\begin{proposition}
If $\mathbf{H}^{(t)}$ is a positive semi-definite matrix for each $t$, then iterates $\bbeta^{(0)}, \bbeta^{(1)}, \bbeta^{(2)}, ...$ produced by a GPX-ECME algorithm (Algorithm \ref{alg:gpx_ecme}) increase the penalized log-likelihood function at each iteration. That is, for any $t \geq 0$,
\begin{equation}
p\ell_{o, \mathbf{s}}^{\boldsymbol{\eta}}(\bbeta^{(t+1)} | \mathbf{y} ) \geq p\ell_{o,\mathbf{s}}^{\boldsymbol{\eta}}(\bbeta^{(t)}|\mathbf{y}). \nonumber
\end{equation}
\end{proposition}

The main purpose for considering surrogate Q-functions of the form (\ref{eq:surrogateQ_form}) in a GPX-ECME algorithm is to allow for more easily-computable parameter updates. 
A choice of $\mathbf{H}^{(t)}$ which typically ensures an easy-to-compute maximizer of $\tilde{Q}_{PX}^{\boldsymbol{\eta}}(\bbeta,1|\bbeta^{(t)}, \alpha_{0})$ is $\mathbf{H}^{(t)} = \mathbf{D}^{(t)} - \mathbf{X}^{T}\mathbf{S}\mathbf{W}(\bbeta^{(t)})\mathbf{X}$, where $\mathbf{D}^{(t)}$ is a $p \times p$ diagonal matrix with nonnegative diagonal entries $d_{jj}^{(t)}$. 
When using a matrix $\mathbf{H}^{(t)}$ of this form, the surrogate Q-function (\ref{eq:surrogateQ_form}) with arguments $(\bbeta, 1)$ reduces to
\begin{equation}
\tilde{Q}_{PX}^{\boldsymbol{\eta}}(\bbeta, 1|\bbeta^{(t)}, \alpha_{0})
= \bbeta^{T}[\mathbf{X}^{T}\mathbf{S}\mathbf{u} + \mathbf{a}_{t}] - \frac{1}{2}\bbeta^{T}\mathbf{D}^{(t)}\bbeta - P_{\boldsymbol{\eta}}(\bbeta) + C, 
\label{eq:qfn_pgd}
\end{equation}
where $\mathbf{a}_{t} = [\mathbf{D}^{(t)} - \mathbf{X}^{T}\mathbf{S}\mathbf{W}(\bbeta^{(t)})\mathbf{X}]\bbeta^{(t)}$. 
Because $\mathbf{D}^{(t)}$ is diagonal, maximizing (\ref{eq:qfn_pgd}) is often straightforward as one can maximize each component $\beta_{j}$ of $\tilde{Q}_{PX}^{\boldsymbol{\eta}}(\bbeta, 1|\bbeta^{(t)}, \alpha_{0})$ separately by 
maximizing a quadratic function plus the $j^{th}$ component of the penalty function $P_{\boldsymbol{\eta}, j}(\beta_{j})$.

\begin{algorithm}[H]
\setstretch{1.25}
\caption{(GPX-ECME for Penalized Logistic Regression). 
In the description of the algorithm,
$\mathbf{u} = (u_{1}, \ldots, u_{n})^{T}$, where $u_{i} = y_{i} - m_{i}/2$.}
\begin{algorithmic}[1]
\State Given $\bbeta^{(0)} \in \mathbb{R}^{p}$.
\For{t=0,1,2,... until convergence}
\State Compute the $n \times n$ diagonal weight matrix $\mathbf{W}(\bbeta^{(t)})$ whose
$i^{th}$ diagonal element is
\begin{equation}
\omega(\mathbf{x}_{i}^{T}\bbeta, m_{i}) = \frac{m_{i}}{2\mathbf{x}_{i}^{T}\bbeta^{(t)}}\tanh\big( \mathbf{x}_{i}^{T}\bbeta^{(t)}/2 \big). \nonumber
\end{equation}
\State Compute $\bbeta^{(t+1), GEM}$:
\vspace{-.2cm}
\begin{equation}
\bbeta^{(t+1), GEM} = \argmin_{\boldsymbol{\beta} \in \mathbb{R}^{p}} \Big[ \frac{1}{2}\bbeta^{T}\mathbf{X}^{T}\mathbf{S}\mathbf{W}(\bbeta^{(t)})\mathbf{X}\bbeta - \bbeta^{T}\mathbf{X}^{T}\mathbf{S}\mathbf{u} + \frac{1}{2}(\bbeta - \bbeta^{(t)})^{T}\mathbf{H}^{(t)}(\bbeta - \bbeta^{(t)}) + P_{\boldsymbol{\eta}}( \bbeta ) \Big]. \nonumber
\vspace{-.2cm}
\end{equation}
\State Compute
\begin{equation}
\rho^{(t+1)} = \argmax_{\rho \in \mathbb{R}} \Bigg[ p\ell_{o, \mathbf{s}}^{\boldsymbol{\eta}}( \rho \bbeta^{(t+1), GEM}|\mathbf{y} ) \Bigg]. \nonumber
\end{equation}
\State Set $\bbeta^{(t+1)} = \rho^{(t+1)}\bbeta^{(t+1), GEM}$.
\EndFor
\end{algorithmic}
\label{alg:gpx_ecme}
\end{algorithm}

If choosing $\mathbf{H}^{(t)} = \mathbf{D}^{(t)} - \mathbf{X}^{T}\mathbf{W}(\bbeta^{(t)})\mathbf{X}$ and setting all the diagonal elements of $\mathbf{D}^{(t)}$ equal so that $\mathbf{D}^{(t)} = \kappa_{t}^{-1}\mathbf{I}_{p}$ for some $\kappa_{t} > 0$, $\mathbf{H}^{(t)}$ will be positive semi-definite provided that
$\kappa_{t}^{-1} \geq \lambda_{max}\big( \mathbf{X}^{T}\mathbf{S}\mathbf{W}(\bbeta^{(t)})\mathbf{X} \big)$, where 
$\lambda_{max}\big( \mathbf{X}^{T}\mathbf{S}\mathbf{W}(\bbeta^{(t)})\mathbf{X}\big)$ is the maximum eigenvalue of $\mathbf{X}^{T}\mathbf{S}\mathbf{W}(\bbeta^{(t)})\mathbf{X}$. When using $\mathbf{D}^{(t)} = \kappa_{t}^{-1}\mathbf{I}_{p}$, it follows from (\ref{eq:px_Qfunction}) that
the term $\bbeta^{(t+1),GEM}$ computed in Step 4 of Algorithm \ref{alg:gpx_ecme} can be be expressed as
\begin{eqnarray}
\bbeta^{(t+1), GEM} 
&=& \argmin_{\boldsymbol{\beta} \in \mathbb{R}^{p}} \Big[ \frac{1}{2}\bbeta^{T}\bbeta -\kappa_{t}\bbeta^{T}[\mathbf{X}^{T}\mathbf{S}\mathbf{u} + \mathbf{a}_{t}]  + \kappa_{t}P_{\boldsymbol{\eta}}(\bbeta) \Big]. 
\label{eq:gem_update_kappa}
\end{eqnarray}
Each component $\beta_{j}^{(t+1), GEM}$ of $\bbeta^{(t+1), GEM}$ can be updated independently of the other components via
\begin{equation}
\beta_{j}^{(t+1), GEM} = \argmin_{\beta_{j} \in \mathbb{R}} \Bigg[ \frac{\beta_{j}^{2}}{\kappa_{t}}  - 2\Big(\sum_{i=1}^{n} x_{ij}s_{i}u_{i} + a_{j,t}\Big)\beta_{j} + 2P_{\boldsymbol{\eta}, j}(\beta_{j}) \Bigg], 
\label{eq:gpxecme_component_update}
\end{equation}
where $a_{j, t}$ denotes the $j^{th}$ component of $\mathbf{a}_{t}$.

The updates (\ref{eq:gpxecme_component_update}) have a closed form for a number of common penalty functions. For example, in the case of the elastic net penalty function where $P_{\boldsymbol{\eta}, j}(\beta_{j}) = \lambda_{1}|\beta_{j}| + \tfrac{\lambda_{2}}{2}\beta_{j}^{2}$, the update for the $j^{th}$ component of $\bbeta^{(t)}$ is given by
\begin{equation}
\beta_{j}^{(t+1), GEM}
= \begin{cases}
\frac{1}{1/\kappa_{t} + \lambda_{2}}\Big( \sum_{i=1}^{n} x_{ij}s_{i}u_{i} + a_{j,t} - \lambda_{1} \Big) & \textrm{ if } \sum_{i=1}^{n} x_{ij}s_{i}u_{i} + a_{j,t} > \lambda_{1} \nonumber \\
\frac{1}{1/\kappa_{t} + \lambda_{2}}\Big( \sum_{i=1}^{n} x_{ij}s_{i}u_{i} + a_{j,t} + \lambda_{1} \Big) & \textrm{ if } \sum_{i=1}^{n} x_{ij}s_{i}u_{i} + a_{j,t} < -\lambda_{1} \nonumber \\
0 & \text{ if } \Big| \sum_{i=1}^{n} x_{ij}s_{i}u_{i} + a_{j,t} \Big| \leq \lambda_{1} \nonumber
\end{cases}
\end{equation}
While other popular choices of the penalty function such as the  smoothly clipped absolute deviation (SCAD) (\cite{fan2001}) penalty may not necessarily yield direct, closed-form solutions, the parameter updates may be directly found by performing a univariate minimization separately for each component of the parameter vector. 

\subsection{Connections to Proximal Gradient Descent} \label{ss:prox_grad}

Proximal gradient methods (\cite{Parikh2014}) are commonly used to optimize a function which can be expressed as the sum of a continuously differentiable function and a non-smooth function. In the context of penalized logistic regression where the aim is to minimize the composite function $p\ell_{o, \mathbf{s}}^{\boldsymbol{\eta}}(\bbeta|\mathbf{y}) = -\ell_{o, \mathbf{s}}(\bbeta| \mathbf{y}) + P_{\boldsymbol{\eta}}(\bbeta)$, the proximal gradient descent update with steplength $\kappa_{t}$ of a current parameter estimate $\bbeta^{(t)}$ is given by
\begin{equation}
\bbeta^{(t+1)} = \textrm{prox}_{\kappa_{t}P_{\boldsymbol{\eta}}}\Big( \bbeta^{(t)} + \kappa_{t}\nabla \ell_{o, \mathbf{s}}(\bbeta^{(t)}|\mathbf{y}) \Big),
\label{eq:pgd_definition}
\end{equation}
where the proximal operator $\textrm{prox}_{\kappa_{t}P_{\boldsymbol{\eta}}}:\mathbb{R}^{p} \longrightarrow \mathbb{R}^{p}$ of the function $\kappa_{t}P_{\boldsymbol{\eta}}$ is defined as
\begin{equation}
\textrm{prox}_{\kappa_{t}P_{\boldsymbol{\eta}}}( \bgamma ) = \argmin_{\boldsymbol{\beta}}\Big[ \frac{1}{2}\bbeta^{T}\bbeta -\bbeta^{T}\bgamma + \kappa_{t}P_{\eta}(\bbeta)  \Big]. \nonumber
\end{equation}
To elaborate on the connection between the proximal gradient update and the GPX-ECME procedure, we first note that the gradient of the observed-data log-likelihood function (\ref{eq:observed_loglik})
is $\nabla \ell_{o, \mathbf{s}}(\bbeta|\mathbf{y}) = \mathbf{X}^{T}\mathbf{S}\{ \mathbf{y} - \mu(\mathbf{X}\bbeta) \}$,
where  $\mu(\mathbf{X}\bbeta)$ is an $n \times 1$ vector whose $i^{th}$ component $[\mu(\mathbf{X}\bbeta)]_{i}$
is given by $[\mu(\mathbf{X}\bbeta)]_{i} = m_{i}\textrm{expit}(\mathbf{x}_{i}^{T}\bbeta)$.
Using the connection (\ref{eq:meanfn_connection}) between the mean function $\mu( \mathbf{X}\bbeta )$ and the weight matrix 
$\mathbf{W}( \bbeta )$, we can write the gradient of $\ell_{o, \mathbf{s}}(\bbeta|\mathbf{y})$ as
\begin{equation}
\nabla \ell_{o, \mathbf{s}}(\bbeta|\mathbf{y}) = \mathbf{X}^{T}\mathbf{S} \{ \mathbf{u} - \mathbf{W}(\bbeta)\mathbf{X}\bbeta \}. \nonumber
\end{equation}
Thus, the proximal gradient update (\ref{eq:pgd_definition}) for penalized logistic can be expressed as 
\begin{eqnarray}
\bbeta^{(t+1), PGD} &=& \textrm{prox}_{\kappa_{t}P_{\boldsymbol{\eta}}}\Big( \bbeta^{(t)} + \kappa_{t}\mathbf{X}^{T}\mathbf{S}\{ \mathbf{u} - \mathbf{W}(\bbeta^{(t)})\mathbf{X}\bbeta^{(t)} \} \Big) \nonumber \\
&=& \argmin_{\boldsymbol{\beta} \in \mathbb{R}^{p}}\Big[ \frac{1}{2}\bbeta^{T}\bbeta - \bbeta^{T}\Big(\bbeta^{(t)} + \kappa_{t}\mathbf{X}^{T}\mathbf{S}\{ \mathbf{u} - \mathbf{W}(\bbeta^{(t)})\mathbf{X}\bbeta^{(t)} \}\Big) + \kappa_{t}P_{\boldsymbol{\eta}}(\bbeta) \Big]. 
\label{eq:pgd_update}
\end{eqnarray}

It is interesting to note that the proximal gradient descent update (\ref{eq:pgd_update}) is exactly the same as the 
update $\bbeta^{(t+1),GEM}$ in (\ref{eq:gem_update_kappa}), which is part of a GPX-ECME algorithm with $\mathbf{H}^{(t)} = \kappa_{t}^{-1}\mathbf{I}_{p} - \mathbf{X}^{T}\mathbf{S}\mathbf{W}(\bbeta^{(t)})\mathbf{X}$. 
In other words, the GPX-ECME algorithm with this choice of $\mathbf{H}^{(t)}$ may be interpreted as a proximal gradient 
descent algorithm where one first takes a proximal gradient descent step with steplength $\kappa_{t}$ and then multiplies
this update by the optimal scalar $\rho^{(t+1)}$ which is optimal in the sense that it leads to the largest increase in the 
observed-data penalized log-likelihood.

\subsection{Connection to an MM algorithm for Logistic Regression} \label{ss:mm_alg}
If one chooses $\mathbf{H}^{(t)}$ to be $\mathbf{H}^{(t)} = \mathbf{X}^{T}\mathbf{S}[\kappa_{t}\mathbf{I}_{p} - \mathbf{W}(\bbeta^{(t)})]\mathbf{X}$ for a positive scalar $\kappa_{t} > 0$ in a GPX-ECME algorithm, the 
surrogate Q-function becomes
\begin{equation}
\tilde{Q}_{PX}^{\boldsymbol{\eta}}(\bbeta, 1|\bbeta^{(t)}, \alpha_{0})
= \bbeta^{T} \mathbf{X}^{T}\mathbf{S}\mathbf{b}^{(t)}(\kappa_{t}) - \frac{\kappa_{t}}{2}\bbeta^{T}\mathbf{X}^{T}\mathbf{S}\mathbf{X}\bbeta - 
P_{\boldsymbol{\eta}}(\bbeta) + C,
\end{equation}
where $\mathbf{b}^{(t)}(\kappa_{t}) = \mathbf{u} + [\kappa_{t}\mathbf{I}_{p} - \mathbf{W}(\bbeta^{(t)})]\mathbf{X}\bbeta^{(t)}$ and $C$ is a constant independent of $\bbeta$. In this context, the matrix $\mathbf{H}^{(t)}$ is guaranteed to be positive semi-definite if $\kappa_{t}$ greater than or equal to the largest diagonal element of $\mathbf{W}( \bbeta^{(t)} )$. For example, setting $\kappa_{t} = \kappa_{t}^{*}$ where
\begin{equation}
\kappa_{t}^{*} = \max\Big\{ \omega(\mathbf{x}_{1}^{T}\bbeta^{(t)}, m_{1}), \ldots, \omega(\mathbf{x}_{n}^{T}\bbeta^{(t)}, m_{n}) \Big \} \nonumber 
\end{equation}
guarantees that $\mathbf{H}^{(t)}$ is positive semi-definite. When there is no penalty (i.e., $P_{\boldsymbol{\eta}}(\bbeta) = 0$), maximizing 
$\tilde{Q}_{PX}^{\boldsymbol{\eta}}(\bbeta, 1|\bbeta^{(t)}, \alpha_{0})$ with respect to $\bbeta$
leads to the following formula for $\bbeta^{(t+1), GEM}$
\begin{equation}
\bbeta^{(t+1), GEM} = 
\argmax_{\boldsymbol{\beta} \in \mathbb{R}^{p}} \tilde{Q}_{PX}^{\boldsymbol{\eta}}(\bbeta, 1|\bbeta^{(t)}, \alpha_{0})
= \frac{1}{\kappa_{t}}( \mathbf{X}^{T}\mathbf{S}\mathbf{X})^{-1}\mathbf{X}^{T}\mathbf{S}\mathbf{b}^{(t)}( \kappa_{t}).
\label{eq:pxmm_update}
\end{equation}
The update $\bbeta^{(t+1)}$ is then found by finding $\rho^{(t+1)} = \argmax_{\rho \in \mathbb{R}}p\ell_{o, \mathbf{s}}( \rho\bbeta^{(t+1), GEM}| \mathbf{y})$ and setting $\bbeta^{(t+1)} = \rho^{(t+1)}\bbeta^{(t+1), GEM}$.

The algorithm defined by the update (\ref{eq:pxmm_update}) is closely related to the MM algorithm for logistic regression described
in \cite{bohning1988} and also detailed in \cite{hunter2004}. To see why, if we use the fact that (\ref{eq:meanfn_connection}) implies 
\begin{equation}
\mathbf{b}^{(t)}( \kappa_{t}) = \mathbf{u} + \big[ \kappa_{t}\mathbf{I} - \mathbf{W}(\bbeta^{(t)}) \big] \mathbf{X}\bbeta^{(t)}
= \kappa_{t}\mathbf{X}\bbeta^{(t)} + \mathbf{y} - \mu(\mathbf{X}^{T}\bbeta^{(t)}), \nonumber
\end{equation}
then (\ref{eq:pxmm_update}) can be rewritten as 
\begin{eqnarray}
\bbeta^{(t+1), GEM} 
&=& \bbeta^{(t)} + \frac{1}{\kappa_{t}}( \mathbf{X}^{T}\mathbf{S}\mathbf{X})^{-1}\mathbf{X}^{T}\mathbf{S}\big\{ \mathbf{y} - \mu(\mathbf{X}^{T}\bbeta^{(t)}) \big\}.
\label{eq:regularmm_update}
\end{eqnarray}
If we were to set $\kappa_{t} = 1/4$ for each $t$ in (\ref{eq:regularmm_update}), then in the case of $m_{1} = m_{2} = \ldots = m_{n} = 1$, then (\ref{eq:regularmm_update}) would correspond exactly to the MM algorithm for logistic regression
update described by \cite{bohning1988}. Note that, in the case when all $m_{i}$ equal $1$, setting $\kappa_{t} = 1/4$ guarantees that $\mathbf{H}^{(t)}$ is positive semi-definite because $\omega(\mathbf{x}^{T}\bbeta, 1) \leq 1/4$ for any value of $\mathbf{x}^{T}\bbeta$. One advantage of using $\kappa_{t} = 1/4$ rather than $\kappa_{t} = \kappa_{t}^{*}$ is that one does not need to compute the maximum of the weights in each iteration; however, this could result in taking slightly smaller ``steps" when compared with using $\kappa_{t}^{*}$.

In the case of an $L_{2}$ penalty where $P_{\boldsymbol{\eta}}(\bbeta) = \tfrac{\lambda}{2}\sum_{j=1}^{p} \beta_{j}^{2}$, 
one would, in (\ref{eq:regularmm_update}), simply replace $(\mathbf{X}^{T}\mathbf{S}\mathbf{X})^{-1}$ with the
matrix $(\mathbf{X}^{T}\mathbf{S}\mathbf{X} + \tfrac{\lambda}{\kappa_{t}}\mathbf{I}_{p})^{-1}$. This
approach for the $L_{2}$-penalized logistic regression problem, which we label the ``PX-MM" algorithm is
summarized in Algorithm \ref{alg:pxmm}. 

\begin{algorithm}[H]
\setstretch{1.25}
\caption{(PX-MM algorithm for Logistic Regression with $L_{2}$ penalty $P_{\boldsymbol{\eta}}(\bbeta) = \frac{\lambda}{2}\bbeta^{T}\bbeta$). 
In the description of the algorithm,
$\mathbf{u} = (u_{1}, \ldots, u_{n})^{T}$, where $u_{i} = y_{i} - m_{i}/2$.}
\begin{algorithmic}[1]
\State Given $\bbeta^{(0)} \in \mathbb{R}^{p}$.
\For{t=0,1,2,... until convergence}
\State Compute the $n \times n$ diagonal weight matrix $\mathbf{W}(\bbeta^{(t)})$ whose
$i^{th}$ diagonal element is
\begin{equation}
\omega(\mathbf{x}_{i}^{T}\bbeta, m_{i}) = \frac{m_{i}}{2\mathbf{x}_{i}^{T}\bbeta^{(t)}}\tanh\big( \mathbf{x}_{i}^{T}\bbeta^{(t)}/2 \big). \nonumber
\end{equation}
\State Set $\kappa_{t}^{*} = \max\big\{ \omega(\mathbf{x}_{1}^{T}\bbeta^{(t)}, m_{1}), \ldots, \omega(\mathbf{x}_{n}^{T}\bbeta^{(t)}, m_{n}) \big \}$.
\State Compute $\bbeta^{(t+1), GEM}$:
\vspace{-.2cm}
\begin{equation}
\bbeta^{(t+1), GEM} = \big( \mathbf{X}^{T}\mathbf{S}\mathbf{X} + \tfrac{\lambda}{\kappa_{t}^{*}}\mathbf{I}_{p} \big)^{-1}\mathbf{X}^{T}\mathbf{S}\mathbf{X}\bbeta^{(t)}
+ \frac{1}{\kappa_{t}^{*}}\big( \mathbf{X}^{T}\mathbf{S}\mathbf{X} + \tfrac{\lambda}{\kappa_{t}^{*}}\mathbf{I}_{p} \big)^{-1}
\mathbf{X}^{T}\mathbf{S}\big\{ \mathbf{y} - \mu(\mathbf{X}^{T}\bbeta^{(t)}) \big\}. \nonumber
\vspace{-.2cm}
\end{equation}
\State Compute
\begin{equation}
\rho^{(t+1)} = \argmax_{\rho \in \mathbb{R}} \Big[ p\ell_{o,\mathbf{s}}^{\boldsymbol{\eta}}( \rho \bbeta^{(t+1), GEM}|\mathbf{y}) \Big]. \nonumber
\end{equation}
\State Set $\bbeta^{(t+1)} = \rho^{(t+1)}\bbeta^{(t+1), GEM}$.
\EndFor
\end{algorithmic}
\label{alg:pxmm}
\end{algorithm}

\section{PX-ECME and Coordinate Descent} \label{sec:coord_descent}
Coordinate descent is often used in the context of $L_{1}$-penalized regression (\cite{Friedman2007}) and $L_{1}$-penalized generalized linear models (\cite{Friedman2010}). Coordinate descent proceeds by updating each regression coefficient one at a time while holding the remaining coefficients fixed. This is frequently an efficient approach as the parameter updates usually have a simple, closed form, and many of the parameters do not change after being mapped to zero.

One could also consider a coordinate descent version of the PX-ECME algorithm where one instead updates the elements of the vector $\btheta$ one at a time rather than in a single step.
To be more specific, consider the parameter-expanded Q-function defined in (\ref{eq:px_qfunction}) with the ``elastic net" penalty $P_{\boldsymbol{\eta},j}(\beta_{j}) = \lambda_{1}|\beta_{j}| + \frac{\lambda_{2}}{2}\beta_{j}^{2}$ and where the first $j-1$ components of $\btheta$ have already been updated
\begin{eqnarray}
& & Q_{PX}^{\lambda}(\theta_{1}^{(t+1)}, \ldots, \theta_{j-1}^{(t+1)}, \theta_{j}, \theta_{j+1}^{(t)}, \ldots, \theta_{p}^{(t)}, \alpha^{(t)}|\btheta^{(t)}, \alpha^{(t)}) = \alpha^{(t)}\theta_{j}\Big( \sum_{i=1}^{n} x_{ij}s_{i}u_{i} \Big) - \nonumber \\
&-& \frac{(\alpha^{(t)})^{2}}{2}\Big( \theta_{j}^{2}\Big\{ \sum_{i=1}^{n}(A_{ij}^{(t)})^{2} + \lambda_{2} \Big\} + 2\theta_{j}\sum_{i=1}^{n} A_{ij}^{(t)} \sum_{k \neq j}\theta_{k}^{(t+j/p)}A_{ik}^{(t)}  \Big) - |\alpha^{(t)}|\lambda_{1}|\theta_{j}| + \tilde{C},
\label{eq:coorddesc_qfn}
\end{eqnarray}
where $A_{ij}^{(t)}$ denotes the $(i,j)$ element of $\mathbf{S}^{1/2}\mathbf{W}^{1/2}(\bbeta^{(t)})\mathbf{X}$, $\tilde{C}$ is a constant not depending on $\theta_{j}$. Maximizing (\ref{eq:coorddesc_qfn}) with respect to $\theta_{j}$ leads to an update of $\theta_{j}^{(t)}$ which has the form
\begin{equation}
\theta_{j}^{(t+1)} = \textrm{sign}\Big( \tfrac{1}{\alpha^{(t)}}V_{j}(\bbeta^{(t)}) - U_{j}(\bbeta^{(t)}, \btheta_{-j}^{(t + j/p)})  \Big)
\max\Big\{ \Big| \tfrac{1}{\alpha^{(t)}}V_{j}(\bbeta^{(t)}) - U_{j}(\bbeta^{(t)}, \btheta_{-j}^{(t + j/p)}) \Big| - \tfrac{1}{\alpha^{(t)}}\tilde{\lambda}(\bbeta^{(t)}), 0 \Big\}.
\label{eq:coorddesc_update}
\end{equation}
Expressions for the terms $V_{j}(\bbeta^{(t)})$, $U_{j}(\bbeta^{(t)}, \btheta_{-j}^{(t + j/p)})$, and $\tilde{\lambda}(\bbeta^{(t)})$ are given in Appendix C.
After completing one ``cycle" of coordinate descent where all $p$ components of $\btheta$ are updated using (\ref{eq:coorddesc_update}), one can
update the value of $\alpha^{(t)}$ by maximizing $p\ell_{o,\mathbf{s}}^{\boldsymbol{\eta}}(\alpha\btheta| \mathbf{y})$ with respect to $\alpha$.
After completing this cycle, the regression coefficients would be updated 
via $\bbeta^{(t+1)} = \alpha^{(t+1)}\btheta^{(t+1)}$.

Rather than waiting an entire cycle to update $\alpha^{(t)}$ and the weight matrix $\mathbf{W}(\bbeta^{(t)})$, an alternative is to perform this updating after a ``block" of $k$ of the components (for $k \leq p$) of $\btheta^{(t)}$ have been updated rather than all $p$ components. This can often reduce the total number of coordinate cycles required to converge; however, this comes at the expense of a greater computational cost to perform each coordinate descent cycle. The more general PX-ECME algorithm which updates the value of $\alpha$ and the regression weights after every $k$ coordinate updates is outlined in Algorithm \ref{alg:pxecme_coord_desc} of Appendix C. Note that when $k = p$, this algorithm reduces to the case where $\alpha$ and the weight matrix are only updated after a full cycle of coordinate descent is completed. 

Our experience with running the PX-ECME version of coordinate descent suggests that it typically offers very modest improvements in the total number of iterations compared to coordinate descent with no parameter expansion (i.e., $\alpha^{(t)}=1$ for all $t$), and the extra computational effort involved in updating $\alpha^{(t)}$ often nullifies any
reduction in the total number of coordinate descent iterations. Nevertheless, experimenting with different values of the ``block size" parameter $k$ can have a positive impact in some cases, and hence, 
in some settings, it may be worth exploring different values of $k$ to determine if there are any gains to be had over a ``classic" coordinate descent algorithm which does not have any parameter expansion whatsoever.

\section{Simulations} \label{sec:sims}

\subsection{A Weighted Logistic Regression Example} \label{ss:simple_sim}

In the standard logistic regression setting where all the weights $s_{i}$ are equal, the
usual starting values used for the Newton-Raphson algorithm (i.e., $\bbeta^{(0)} = \mathbf{0}$) are generally quite robust,
and using these starting values with Newton-Raphson produces iterates which converge to the maximum likelihood estimates of the regression coefficients. 
However, in many weighted logistic regression problems where there is large variability in the weights,
the Newton-Raphson algorithm can result in failure even when the starting values of $\bbeta^{(0)} = \mathbf{0}$
are used. To illustrate this, we consider the following 
small numerical example with responses $y_{i}$, scalar covariates $x_{i}$, 
and regression weights $s_{i}$:
\begin{eqnarray}
(y_{1}, \ldots, y_{7}) &=& (1, 0, 1, 1, 1, 0,1) \nonumber \\
(x_{1}, \ldots, x_{7}) &=& (0, 0, 0.001, 100, -1, -1, 0.5) \nonumber \\
(s_{1}, \ldots, s_{7}) &=& (0.4, 0.01, 0.4, 0.01, 0.04, 0.1, 0.04). \label{eq:weighted_example_dat}
\end{eqnarray}

The performance of Newton-Raphson, PX-ECME, and EM using the data and regression weights in (\ref{eq:weighted_example_dat}) and with no penalty term is summarized in Table \ref{tab:weighted_nonconvergence}. We ran each method for 63 iterations because this was the number of iterations
required for PX-ECME to converge when using the stopping criterion $\sqrt{\sum_{j=1}^{p}(\beta_{j}^{(t+1)} - \beta_{j}^{(t)})^{2}} < 10^{-9}$. As shown in Table \ref{tab:weighted_nonconvergence},
the Newton-Raphson procedure diverges even when using the starting values $\bbeta^{(0)} = \mathbf{0}$.
While the first iteration of Newton-Raphson improves the value of the weighted log-likelihood, the subsequent 
iterates begin to quickly diverge. 
In contrast to Newton-Raphson, the PX-ECME algorithm provides more stable monotone convergence to the weighted maximum likelihood estimates
$(\hat{\beta}_{0}, \hat{\beta}_{1}) = (4.39, 5.30)$. While the EM algorithm also enjoys monotone convergence, 
the convergence is considerably slower than PX-EMCE, and it takes $419$ iterations for EM to converge when using the same 
convergence tolerance as PX-ECME. 

\begin{table}[ht]
\centering
\begin{tabular}{c rrcrrcrrr}
\toprule
& \multicolumn{3}{c}{Newton-Raphson} &
  \multicolumn{3}{c}{PX-ECME}  & \multicolumn{3}{c}{EM} \\ 
\cmidrule(r){2-4}\cmidrule(r){5-7}\cmidrule(r){8-10}
k & $\beta_{0}^{(k)}$ & $\beta_{1}^{(k)}$ & $\ell_{o,\mathbf{s}}^{(k)}$ & $\beta_{0}^{(k)}$ & $\beta_{1}^{(k)}$ & $\ell_{o,\mathbf{s}}^{(k)}$  & $\beta_{0}^{(k)}$ & $\beta_{1}^{(k)}$ & $\ell_{o,\mathbf{s}}^{(k)}$ \\
\midrule
 0 & 0.0 & 0.0 & -0.6931 & 0.0 & 0.0 & -0.6931 & 0.0 & 0.0 & -0.6931 \\ 
  1 & 4.26 & 1.97 & -0.2972 & 1.55 & 0.01 & -0.3611 & 1.55 & 0.01 & -0.3611 \\ 
  2 & -79.22 & -10.17 & -80.4768 & 2.09 & 0.02 & -0.3441 & 1.85 & 0.01 & -0.3471 \\ 
  3 & $1.87 \times 10^{16}$ & $4.31 \times 10^{15}$ & $-1.62 \times 10^{15}$ & 2.10 & 0.03 & -0.3432 & 1.97 & 0.02 & -0.3441 \\ 
  4 & $9.60 \times 10^{15}$ & $4.43 \times 10^{15}$ & $-6.13 \times 10^{14}$ & 2.10 & 0.03 & -0.3424 & 2.03 & 0.03 & -0.3429 \\ 
  5 & $4.88 \times 10^{14}$ & $4.56 \times 10^{15}$ & $-1.67 \times 10^{14}$ & 2.10 & 0.05 & -0.3414 & 2.05 & 0.04 & -0.3420 \\ 
  6 & $5.28 \times 10^{14}$ & $4.51 \times 10^{15}$ & $-1.64 \times 10^{14}$ & 2.11 & 0.06 & -0.3404 & 2.07 & 0.05 & -0.3410 \\ 
  7 & $5.68 \times 10^{14}$ & $4.46 \times 10^{15}$ & $-1.61 \times 10^{14}$ & 2.12 & 0.07 & -0.3392 & 2.07 & 0.06 & -0.3400 \\ 
  8 & $6.08 \times 10^{14}$ & $4.42 \times 10^{15}$ & $-1.58 \times 10^{14}$ & 2.13 & 0.09 & -0.3377 & 2.08 & 0.08 & -0.3388 \\ 
  9 & $6.48 \times 10^{14}$ & $4.37 \times 10^{15}$ & $-1.55 \times 10^{14}$ & 2.14 & 0.11 & -0.3360 & 2.08 & 0.09 & -0.3373 \\ 
  10 & $6.88 \times 10^{14}$ & $4.33 \times 10^{15}$ & $-1.52 \times 10^{14}$ & 2.15 & 0.14 & -0.3339 & 2.08 & 0.11 & -0.3357 \\ 
  \vdots & \vdots & \vdots & \vdots & \vdots & \vdots & \vdots & \vdots & \vdots & \vdots \\
  \vdots & \vdots & \vdots & \vdots & \vdots & \vdots & \vdots & \vdots & \vdots & \vdots \\
  63 & $-6.71 \times 10^{15}$ & $2.51 \times 10^{15}$ & $-5.95 \times 10^{15}$ & 4.39 & 5.30 & -0.1376 & 4.01 & 4.83 & -0.1386 \\ 
\bottomrule
\end{tabular}
\caption{Comparison of Newton-Raphson, PX-ECME, and EM using the data and regression weights provided in (\ref{eq:weighted_example_dat})
and with initial values of $(\beta_{0}^{(0)}, \beta_{1}^{(0)}) = (0.0, 0.0)$ for all methods. Values of the regression coefficients $\beta_{0}^{(k)}, \beta_{1}^{(k)}$ and the associated weighted log-likelihood values
$\ell_{o, \mathbf{s}}^{(k)}$ are shown for each method and iterations $k = 0, 1, ..., 10$ and $k = 63$. 
Iterates from the Newton-Raphson algorithm do not converge
while both PX-ECME and EM converge to the maximum likelihood estimates of the regression coefficients albeit with
notably slower convergence from the EM algorithm.} 
\label{tab:weighted_nonconvergence}
\end{table}

\subsection{The Kyphosis Data} \label{ss:kyphosis}
In this simulation study, we utilized the Kyphosis data described in \cite{chambers1992} (pg. 200) and also considered in \cite{Liu1998}.
This dataset contains 81 observations from children who had undergone a corrective spinal surgery. 
The binary outcome in this dataset indicates whether or not kyphosis was present after the operation, and the following additional 3 covariates were also recorded: the age of the child in months, the number of vertebrae, and the number of the topmost vertebra involved in the surgery. 
For this simulation study, we did not use the observed binary outcomes but instead only used the observed covariates and simulated binary responses. Similar to the simulation described in \cite{Liu1998}, we generated pseudo-binary outcomes $y_{i}$ assuming
$P(y_{i} = 1) = 1/[1 + \exp(-3x_{i,num} + x_{i,start})]$, where $x_{i,num}$  and $x_{i,start}$ denote the number of vertebrae and the number of topmost vertebra for the $i^{th}$ child respectively. We generated $500$ datasets in this manner so that each dataset contained the 3 covariate vectors in the original Kyphosis data and the vector of simulated pseudo-outcomes. When estimating the parameters of this model, we did not include a penalty term, and we included an intercept term so that, in total, four regression coefficients needed to be estimated.

\begin{table}[ht]
\centering
\ra{1.1}
\begin{tabular}{l rrr rrr r}
\toprule
\multirow{2}{*}{Method} & \multicolumn{3}{c}{Number of iterations} &
  \multicolumn{3}{c}{Timing} &
  \multicolumn{1}{c}{$\log L(\hat{\theta}) $} \\
\cmidrule(r){2-4}\cmidrule(r){5-7}\cmidrule(r){8-8}
 & median & mean & std. dev. & median & mean & std. dev. & mean \\
\midrule
EM & 424 & 756.96 & 1099.89 & 0.0215 & 0.0410 & 0.0671 & -10.524639 \\ 
  PX-ECME & 49 & 51.61 & 19.43 & 0.0050 & 0.0061 & 0.0058 & -10.524639 \\
  MM & 2242 & 7079.57 & 18123.52 & 0.0765 & 0.2577 & 0.6738 & -10.524639 \\ 
  PX-MM & 139 & 177.88 & 140.69 & 0.0130 & 0.0181 & 0.0170 & -10.524639 \\ 
  NR & 10 & 10.14 & 1.14 & 0.0010 & 0.0015 & 0.0014 & -10.524639 \\ 
  AA1 & 59 & 71.53 & 49.86 & 0.0040 & 0.0050 & 0.0036 & -10.524639 \\
\bottomrule
\end{tabular}
\caption{Simulation results based on the Kyphosis data with simulated outcomes. Median number of iterations required to converge and median timing across simulation replications are presented. The mean value of the log-likelihood at convergence is also shown. Methods evaluated
include the EM algorithm, PX-ECME algorithm, the MM algorithm described in Section \ref{ss:mm_alg}, 
the parameter expanded MM (PX-MM) algorithm (Algorithm 3),
the Newton-Raphson (NR) algorithm, and the order-1 Anderson acceleration of EM described in Section \ref{ss:aa_overview}. }
\label{tab:kyphosis_results}
\end{table}

Using these 500 simulated datasets, we compared the following methods for computing the four regression coefficient estimates: the EM algorithm, Newton-Raphson, the PX-ECME algorithm (Algorithm \ref{alg:basic_pxecme}), the MM algorithm of Section \ref{ss:mm_alg}, the PX-MM algorithm (Algorithm \ref{alg:pxmm}), and the order-1 Anderson acceleration of EM (AA1).
For each method, all regression coefficients were initialized at $0$, and the stopping criteria of $\sqrt{\sum_{j=1}^{p} (\beta_{j}^{(t+1)} - \beta_{j}^{(t)})^{2}} < 10^{-7}$ was used for each method. 

The performance of these procedures is summarized in Table \ref{tab:kyphosis_results}. As shown in Table \ref{tab:kyphosis_results}, the Newton-Raphson procedure performed the best requiring only a median of $10$
Newton-Raphson iterations to converge and a median of $0.001$ seconds to achieve convergence.
The PX-ECME was the second-fastest algorithm in terms of
number of iterations only requiring a median of 49 iterations for 
convergence. Compared to the EM algorithm, this was a roughly ten-fold reduction in the median number of iterations.
PX-ECME and AA1 were quite close in performance with AA1 requiring
a few more iterations than PX-EMCE while having better time performance
than PX-ECME due to a mildly cheaper per iteration cost than
PX-ECME. A notable result from the Kyphosis data simulations was that PX-MM was substantially faster than the MM algorithm. Compared to the MM algorithm, PX-MM had a nearly 17-fold reduction in the median number of iterations and a nearly 
six-fold reduction in the median time to converge.

\subsection{Simulated Outcomes with Autocorrelated Covariates} \label{ss:ar1_sims}

For this simulation study, we generated the elements $x_{i,j}$ of an $n \times p$ design matrix $\mathbf{X}$ using an autoregressive process of order 1 with autocorrelation parameter $\rho \geq 0$. Specifically, for each $i$, the $x_{i,j}$ were generated as $x_{i,1} = 1$, and
\begin{equation}
x_{i,j} = \rho x_{i,j-1} + \varepsilon_{j}, \qquad \textrm{ for } j = 3, \ldots,p, \nonumber
\end{equation}
where $x_{i,2} \sim \textrm{Normal}(0, 1)$ and $\varepsilon_{j} \sim \textrm{Normal}(0, 1)$. 
This simulation design implies the correlation between $x_{i,j}$ and $x_{i,j-1}$ is $\rho$ for any $j \geq 3$. The regression coefficients $\beta_{1},\ldots,\beta_{p}$ were generated independently as
$\beta_{j} = Z_{j}T_{j}$, where $Z_{j} \sim \textrm{Bernoulli}(0.75)$ and $T_{j}$ follows a t distribution with $3$ degrees of freedom. Given the generated covariate vectors $\mathbf{x}_{1}, \ldots, \mathbf{x}_{n}$ and regression coefficient vector $\bbeta$, the responses $y_{i}$ were then generated independently as $y_{i} \sim \textrm{Bernoulli}\big( \{ 1 + \exp(-\mathbf{x}_{i}^{T}\bbeta) \}^{-1} \big)$.

We considered two choices for the number of observations: $n \in \{ 500, 2000\}$, three settings for the number of covariates: $p \in \{5, 50, 200\}$, and three values for the autocorrelation parameter $\rho \in \{0, 0.9, 0.99 \}$. For each of the $18$ simulation settings (i.e., each setting of $n$, $p$, and $\rho$), we ran each procedure for computing the maximum likelihood estimate of $\bbeta$ with no penalty term on $50$ simulated datasets. For all methods considered, we used a maximum of $100,000$ iterations.

Table \ref{tab:ar_results_zerostart} shows summary performance measures of $8$ methods for estimating the logistic regression coefficients in these simulation scenarios. With the exception of Newton-Raphson, each procedure displayed in Table \ref{tab:ar_results_zerostart} is a monotone algorithm as the steplength in gradient descent and a GPX-ECME algorithm (Algorithm 2) were chosen to guarantee an increase in the log-likelihood at each iteration. The GPX-ECME algorithm used here chooses $\mathbf{H}^{(t)} = \kappa_{t}^{-1}\mathbf{I}_{p} - \mathbf{X}^{T}\mathbf{S}\mathbf{W}(\bbeta^{(t)})\mathbf{X}$ (where $\kappa_{t}^{-1}$ is the maximum eigenvalue of $\mathbf{X}^{T}\mathbf{S}\mathbf{W}(\bbeta^{(t)})\mathbf{X}$ ) so that the GPX-ECME update is simply the gradient descent update multiplied by the scalar which maximizies the observed-data log-likelihood.
The summary measures shown in Table \ref{tab:ar_results_zerostart} correspond to performance measures aggregated across all 18 simulation settings. 

For the results shown in the top half of Table \ref{tab:ar_results_zerostart}, the initial value $\bbeta^{(0)}$ was set to the zero vector. While Newton-Raphson has the best overall performance when setting $\bbeta^{(0)} = \mathbf{0}$, PX-ECME shows clear advantages over the EM algorithm and the MM and PX-MM algorithms. 
Indeed, PX-ECME shows a more than ten-fold reduction from EM in the median number of iterations required to converge, and PX-ECME
shows a more than three-fold reduction in the median number of iterations when compared with PX-MM. When compared with EM, PX-ECME also reduced the 
number of cases with very long times to converge which can be observed by noting the even greater reduction in mean number of iterations
for PX-ECME versus EM when compared to the reduction in median iterations for PX-ECME versus EM. While each iteration of PX-ECME requires moderately more computational time than EM, the timing comparisons show that 
the median time to convergence of PX-ECME is nearly 10 times less than
that of the EM algorithm.

The lower half of Table \ref{tab:ar_results_zerostart} shows the performance of each method when the components of the initial vector $\bbeta^{(0)}$
are set by sampling from a standard normal distribution. With these random starting values, all of the methods except Newton-Raphson demonstrated robust performance with all of these methods having very similar performance to the simulation runs where $\bbeta^{(0)} = \mathbf{0}$. Newton-Raphson, however, showed very erratic
performance when the initial values were not set to zero demonstrating this method's considerable sensitivity to the choice of starting values, and in the majority of simulation runs, Newton-Raphson did not converge. 
As with the simulation runs with $\bbeta^{(0)} = \mathbf{0}$,
PX-ECME with random starting values provides a roughly 10-fold
improvement in convergence speed over the EM algorithm.

\begin{table}[ht]
\centering
\ra{1.1}
\begin{tabular}{l rrr rrr r}
\toprule
\multirow{2}{*}{Method} & \multicolumn{3}{c}{Number of iterations} &
  \multicolumn{3}{c}{Timing} &
  \multicolumn{1}{c}{$\log L(\hat{\theta}) $} \\
\cmidrule(r){2-4}\cmidrule(r){5-7}\cmidrule(r){8-8}
 & median & mean & std. dev. & median & mean & std. dev. & mean \\
\midrule
initial values of zero: \\
\cmidrule(r){1-1}
EM & 625 & 4509.12 & 13224.93 & 0.2395 & 5.5697 & 23.0756 & -210.230466 \\ 
  PX-ECME & 48 & 130.98 & 315.16 & 0.0340 & 0.1980 & 0.9346 & -210.230148 \\  
  MM & 3762 & 24038.59 & 36171.98 & 0.3440 & 3.3383 & 6.4779 & -210.277489 \\
  PX-MM & 161 & 1248.01 & 5258.13 & 0.0860 & 0.8922 & 6.1386 & -210.230148 \\
  NR & 10 & 10.51 & 2.92 & 0.0060 & 0.0129 & 0.0180 & -210.230148 \\ 
  GD & 33917 & 48948.32 & 45435.61 & 8.6060 & 44.8929 & 77.4781 & -213.906704 \\ 
  GPX-ECME & 13178 & 38945.89 & 43830.79 & 7.3200 & 54.3209 & 93.0919 & -211.519659 \\ 
  GDBT & 9766 & 39079.08 & 44644.20 & 4.0385 & 44.5298 & 81.2833 & -211.461741 \\ 
  AA1 & 42 & 89.09 & 370.13 & 0.0200 & 0.1109 & 0.8442 & -210.230148 \\
\midrule
random initial values: \\
\cmidrule(r){1-1}
   EM & 596 & 4431.32 & 12871.93 & 0.2480 & 5.6423 & 24.3456 & -203.681932 \\ 
  PX-ECME & 50 & 122.06 & 193.26 & 0.0365 & 0.1649 & 0.3405 & -203.681930 \\  
  MM & 3680 & 24055.37 & 35992.27 & 0.4185 & 3.1970 & 6.0510 & -203.720918 \\
  PX-MM & 158 & 998.80 & 2754.27 & 0.0900 & 0.6194 & 2.1296 & -203.681930 \\
  NR$^{*}$ & 100000 & 87223.14 & 33400.38 & 39.4695 & 120.7834 & 146.0996 & -4.02 $\times 10^{16}$ \\ 
  GD & 31308 & 48637.68 & 45354.11 & 8.8395 & 45.1825 & 78.0428 & -207.656757 \\ 
  GPX-ECME & 13901 & 39070.57 & 43749.47 & 7.6315 & 53.8382 & 91.0420 & -205.777709 \\ 
  GDBT & 9094 & 39178.89 & 44816.03 & 3.9850 & 44.9879 & 81.2179 & -204.791037 \\ 
  AA1 & 46 & 80.60 & 172.71 & 0.0210 & 0.0917 & 0.4262 & -203.681930 \\ 
  \bottomrule
\end{tabular}
\caption{Results for simulated outcomes with autocorrelated covariates. 
PX-MM, NR, and AA1 denote the parameter-expanded MM algorithm (Algorithm 3), 
Newton-Raphson, and order-1 Anderson acceleration respectively.
GD denotes gradient descent, and GDBT denotes gradient 
descent with backtracking. The GPX-ECME algorithm (Algorithm 2) used here 
chooses $\mathbf{H}^{(t)}$ so that the parameter update in each iteration is found by multiplying the gradient descent update by an optimal 
scalar.
Each method was stopped after $100,000$ iterations. \\
$^{*}$Newton-Raphson resulted in numerical errors in $16$ out of $900$ simulation runs; these runs were discarded
when tabulating the summary measures of performance.}
\label{tab:ar_results_zerostart}
\end{table}

\subsection{The Madelon Data} \label{ss:madelon}

In this simulation study, we consider a weighted
logistic regression example with both an $L_{1}$ penalty (i.e, $P_{\eta}(\bbeta) = \eta \sum_{j=1}^{p} |\beta_{j}|$)
and an $L_{2}$ penalty function (i.e., $P_{\eta}(\bbeta) = \frac{\eta}{2}\sum_{j=1}^{p} \beta_{j}^{2}$). We used the Madelon dataset (\cite{guyon2004}) which is
available for download from the UCI machine learning
repository. This dataset has $n = 2600$ observations
and $p = 500$ covariates.
While the same Madelon dataset was used for all
simulation replications, a different set of weights $s_{1}, \ldots, s_{2600}$
were drawn for each of 10 simulation replications. The 
weights $s_{i}$ were sampled independently from an exponential distribution
with rate parameter 1. We generated 10 sets of weights for both the 
$L_{2}$ and $L_{1}$ penalized cases.

For each of the ten sets of weights, we
obtained parameter estimates $\hat{\bbeta}(\eta_{k})$
across a decreasing sequence $\eta_{1} > \eta_{2} > \ldots > \eta_{9}$ of nine tuning parameters. The same values of $\eta_{k}$ were used
for both the $L_{1}$ and $L_{2}$-penalized simulation runs. 
These values were $\eta_{1} = 5000$, $\eta_{2} = 1000$,
$\eta_{3} = 500$, $\eta_{4} = 200$, $\eta_{5} = 50$,
$\eta_{6} = 10$, $\eta_{7} = 2$, $\eta_{2}$, $\eta_{9} = 0.1$.
For a given set of weights, the parameter estimate $\bbeta$ for the previous value of $\eta$ was used as the starting value for the subsequent value of $\eta$. That is, $\hat{\bbeta}(\eta_{k})$ was used as the starting
value for each method when $\eta$ was equal to $\eta_{k+1}$.
The following methods were evaluated for the $L_{1}$-penalized case:
coordinate descent with EM-based weights, coordinate descent with Newton-Raphson-based weights, proximal gradient descent (PGD) with a fixed steplength, parameter-expanded proximal gradient descent (GPX-ECME), and gradient descent with backtracking. The following methods were evaluated for the $L_{2}$-penalized case:
EM, PX-ECME, MM, PX-MM, Newton-Raphson, fixed steplength gradient descent, GPX-ECME, gradient descent with backtracking, and order-1 Anderson acceleration.
For the $L_{1}$-penalized simulations, we set a maximum of $100000$ iterations
for the non-coordinate descent methods and a maximum of $1000$ iterations for the coordinate descent methods (one iteration for this case is a full cycle that updates all the regression coefficients), and for the $L_{2}$-penalized simulations, we set a maximum of $10000$ iterations.

The madelon-data simulation results are shown in Table \ref{tab:madelon_sims}. Note that the summary measures for the number of iterations and timings are aggregated across all $10$ replications and all values of $\eta$ and are not separated by value of $\eta$.
In the $L_{1}$-penalized simulations, the parameter-expanded version (GPX-ECME) of proximal gradient descent (PGD) performs well in terms
of both timing and final value of the objective function.
Specifically, the timing required by GPX-ECME was not much greater than 
PGD while the average value of the objective function achieved by GPX-ECME at the final iteration was much better than that achieved by PGD. While
proximal gradient descent with backtracking achieved better
log-likelihood values after $100,000$ iterations, this was at a cost
of much slower speeds than either PGD or GPX-ECME.

In the $L_{2}$-penalized simulations, Newton-Raphson
had the smallest median number of iterations needed for convergence.
However, of the 90 simulation runs, there were 14 runs where Newton-Raphson diverged which
led to a very large mean time spent and mean number of 
iterations needed. 
In contrast to Newton-Raphson, both EM and MM had more robust performance
without sacrificing much computational speed relative to Newton-Raphson.
Overall, both PX-ECME and PX-MM delivered speed advantages
over the EM and MM algorithms respectively,
but the additional gains in computational speed were,
overall, quite modest. This seems to be an example 
where both EM and MM are quite fast relative to Newton-Raphson
(when Newton-Raphson converges), and hence, it is difficult
for PX-ECME or PX-MM to make substantial improvements.
The strong performance of order-1 Anderson acceleration (AA1) in 
the $L_{2}$-penalized simulations is also interesting to note.
In addition to demonstrating considerable robustness by converging to the maximizer of the objective function in every simulation run, AA1 was somewhat faster than PX-ECME and was even competitive with Newton-Raphson in 
median timing.

\begin{table}[ht]
\centering
\ra{1.1}
\begin{tabular}{l rrr rrr rr}
\toprule
\multirow{2}{*}{Method} & \multicolumn{3}{c}{Number of iterations} &
  \multicolumn{2}{c}{Timing} &
  \multicolumn{1}{c}{$\log L(\hat{\theta}) $} \\
\cmidrule(r){2-4}\cmidrule(r){5-6}\cmidrule(r){7-7}\cmidrule(r){8-8}
 & median & mean & std. dev. & median & mean & mean & not converged \\
\midrule
$L_{1}$ penalty: \\
\cmidrule(r){1-1}
Coord Desc(EM) & 1000 & 1000 & 0.00 & 2212.20 & 2212.06 & -1610.69 & 90 \\  
Coord Desc(NR) & 1000 & 1000 & 0.00 & 2737.15 & 2737.40 & -1580.09 & 90 \\ 
  PGD & 100000 & 73635.87 & 35608.04 & 852.77 & 630.84 & -1573.44 & 49 \\
  GPX-ECME & 100000 & 75781.66 & 33794.31 & 1268.08 & 964.46 & -1515.11 & 51 \\
  PGDBT & 100000 & 72641.41 & 40250.80 & 28457.42 & 20723.50 & -1483.10 & 60 \\ 
\midrule
$L_{2}$ penalty: \\
\cmidrule(r){1-1}
EM & 31 & 30.76 & 4.25 & 9.61 & 9.52 & -1200.41 & 0 \\ 
  PX-ECME & 24 & 23.50 & 3.27 & 7.58 & 7.39 & -1200.41 & 0 \\
  MM & 56 & 55.04 & 9.92 & 0.89 & 0.89 & -1200.41 & 0 \\ 
  PX-MM & 43 & 41.97 & 8.30 & 0.99 & 1.00 & -1200.41 & 0 \\ 
  NR & 9 & 1563.11 & 3641.38 & 5.03 & 874.79 & -8.05 $\times 10^{13}$ & 14 \\ 
  GD & 10000 & 10000 & 0 & 2696.13 & 2697.28 & -1664.13 & 90 \\ 
  GPX-ECME & 10000 & 10000 & 0 & 2735.57 & 2737.46 & -1530.06 & 90 \\ 
  GDBT & 10000 & 10000 & 0 & 2773.56 & 2782.01 & -1597.34 & 90 \\ 
  AA1 & 18 & 18.06 & 2.30 & 5.66 & 5.63 & -1200.41 & 0 \\ 
  \bottomrule
\end{tabular}
\caption{Results from the madelon simulation study. 
Ten different weighted log-likelihood functions were tested using 
$10$ different sets of exponentially distributed weights, and, for each set of weights, each method was run for $9$ different choices of a penalty parameter. Methods shown include coordinate descent with EM or Newton-Raphson (NR) weights, proximal gradient descent (PGD), proximal gradient descent with backtracking (PGDBT), and order-1 Anderson acceleration(AA1).
For each method, the number of simulation runs (out of 90 simulation runs in total) where the method did not converge is also reported.
}
\label{tab:madelon_sims}
\end{table}

\section{Discussion} \label{sec:conc}

In this article, we have proposed and investigated the 
performance of an accelerated version of a P\'olya-Gamma-based EM algorithm for weighted penalized logistic regression. 
Our method provides a useful alternative algorithm
that is stable, efficient, directly implementable, and provides
monotone convergence regardless of the weights or starting 
values chosen by the user.
We established that our PX-ECME algorithm has a convergence rate 
guaranteed to be as fast as the original P\'olya-Gamma EM algorithm, and empirically, across a range of different simulation examples,
our PX-ECME algorithm accelerates the convergence speed roughly by a factor of ten when compared to the EM algorithm. 
We also proposed a generalized version of the P\'olya-Gamma EM algorithm
and explored a PX-ECME version of this generalized EM algorithm. 
This generalization of EM can be useful in certain high dimensional contexts where performing the full M-step in every iteration of the EM algorithm can become computationally expensive.
Our generalized EM algorithm includes both proximal gradient 
descent and the logistic regression MM algorithm as special cases, and
in our simulations, we showed that the parameter-expanded version of 
the MM algorithm can have substantially faster convergence than the corresponding MM algorithm.

While Newton-Raphson typically converges very quickly in cases
where Newton-Raphson is well-behaved, a chief motivation
behind the development of the PX-ECME algorithm was to explore
an alternative optimization procedure which has relatively fast convergence 
yet still has stable, robust performance regardless of the choice
of starting values and choice of weights. Improving the robustness of Newton-Raphson becomes a more salient issue 
in cases where one is optimizing a weighted log-likelihood with highly
variable weights. Indeed, the first example 
in Section \ref{sec:sims} shows a simple case where Newton-Raphson
regularly diverges even when the initial values are
set to the typically robust starting value of zero for all parameters,
and in this case, both EM and PX-ECME converge to the correct
value. In addition to greater robustness and stability,
another important issue is the fact that the EM algorithm framework can 
often be more easily adapted to generate stable, straightforward
parameter updates in more complex variations of a logistic regression model
where there are additional latent variables in the probability model of interest. A common example of this is when one posits an additional probability model for the covariates in order to perform 
parameter estimation in the context of missing data. In Appendix A, we describe
a PX-ECME algorithm for one example of a missing-data model, and we 
demonstrate how the PX-ECME parameter updates for this missing-data model
are very similar to the PX-ECME updates for fully observed covariates.

While PX-ECME is a straightforward modification of EM that substantially
boosts convergence speed, it is certainly possible that 
applying other monotone acceleration schemes to the PX-ECME 
iteration could improve convergence speed even further. 
Specifically, using ``off-the-shelf" acceleration schemes
which directly accelerate the convergence of an iterative sequence
without modifying the ``base" optimization algorithm in any way
could be applied directly to iterates of the PX-ECME algorithm. 
One such approach would be to apply the monotone order-1 Anderson acceleration
to the PX-ECME updates rather than the EM updates as was done in
Section \ref{ss:aa_overview}. This would likely provide further speed gains 
in addition to those of PX-ECME over EM. 
It would also be of interest to explore the performance of 
applying other, higher-order off-the-shelf acceleration schemes to 
iterates of the PX-ECME algorithms such
as the SQUAREM procedure (\cite{Varadhan2008}) or the quasi-Newton acceleration scheme of \cite{Zhou2011}.

The PX-ECME parameter update corresponds to multiplying the EM parameter update by a single scalar, where the scalar is chosen to maximize
the observed log-likelihood. This can be thought of as
finding the parameter update by searching across parameter updates which are a scalar multiple
of the EM update. This type of ``multiplicative
search" was chosen because it fits naturally into the parameter-expansion framework, and because
a similar type of multiplicative update
has been found to work well in the context of accelerating EM algorithms for probit regression. 
Though not explored here, it is likely that alternative approaches based on using the EM algorithm search direction could
deliver relatively similar performance to our PX-ECME algorithm. 
For example, searching across parameter updates that are linear combinations of the previous parameter value and the EM update
could be an effective strategy that has relatively similar performance
to our PX-ECME algorithm.

\medskip
\begin{center}
{\large\bf SUPPLEMENTARY MATERIAL}
\end{center}

\begin{description} 
\item[R-package:] An R-package \verb"pxlogistic" implementing the methods described in the article is available at \url{https://github.com/nchenderson/pxlogistic}
\end{description}

\appendix

\section{A PX-ECME algorithm with Missing Covariates}

As an example of a direct EM algorithm that can incorporate modeling of missing covariates, we consider a setting similar to that described in \cite{ibrahim1990}. Specifically, 
we consider the case where all of the covariates are binary. 
To develop an EM algorithm in this context, we let 
$\mathbf{x}_{ic}$ denote the ``complete version" of the vector $\mathbf{x}_{i}$, where we allow for the possibility that $\mathbf{x}_{i}$ can contain missing values. We can express $\mathbf{x}_{ic} = (\mathbf{x}_{i,obs}, \mathbf{x}_{i,mis})$, where $\mathbf{x}_{i,obs}$ is the collection
of observed values from $\mathbf{x}_{ic}$ and $\mathbf{x}_{i,mis}$
is the collection of missing values from $\mathbf{x}_{ic}$.

\bigskip

\noindent
The assumed joint distribution for the covariate vector $\mathbf{x}_{ic}$ is given by
\begin{equation}
p(\mathbf{x}_{ic}|\bgamma) = \prod_{k=1}^{2^{p-1}} \gamma_{k}^{I_{k}(\mathbf{x}_{ic})}, 
\label{eq:xdistribution}
\end{equation}
where $k = 1, \ldots, 2^{p-1}$ indexes the $2^{p-1}$
possible observed values of $\mathbf{x}_{ic} = (1, \mathbf{z}_{ic}^{T})^{T}$, where $\mathbf{z}_{ic} \in \{0, 1\}^{p-1}$. In (\ref{eq:xdistribution}), $I_{k}(\mathbf{x}_{ic}) = 1$ if
$\mathbf{x}_{ic}$ equals the $k^{th}$ possible observed value of $\mathbf{x}_{ic}$ and equals $0$ otherwise. Note that 
this missing data model is only useful in practice
when $p$ is relatively small.

\bigskip

\noindent
Using similar notation to that used in our main manuscript, we consider the following P\'olya-Gamma representation of the logistic regression model
\begin{eqnarray}
y_{i}|W_{i}, \mathbf{x}_{ic} &\sim& \textrm{Binomial}\Big(m_{i}, \{1 + \exp(-\mathbf{x}_{ic}^{T}\bbeta)\}^{-1} \Big) \nonumber \\
W_{i}|\mathbf{x}_{ic} &\sim& PG(m_{i}, \mathbf{x}_{ic}^{T}\bbeta) \nonumber \\
\mathbf{x}_{ic} &\sim& p(\cdot| \bgamma), \label{eq:full_hierarchy}
\end{eqnarray}
where $p(\cdot|\bgamma)$ refers to the same probability model
stated in (\ref{eq:xdistribution}). We would also assume that the data are missing at random (MAR) so that if $\mathbf{r}_{i}$ denotes the vector of missingness indicators, then
\begin{equation}
p(\mathbf{x}_{ic}|\mathbf{x}_{i,obs}, \mathbf{r}_{i}, \bgamma)
= p(\mathbf{x}_{ic}|\mathbf{x}_{i,obs},\bgamma)
= \frac{p(\mathbf{x}_{ic}|\bgamma)}{\sum_{k} p(\mathbf{x}_{ic}=\mathbf{d}_{k}|\bgamma)I(k \in \mathbf{A}_{i}) }, \nonumber  
\end{equation}
where $\mathbf{x}_{i,obs}$ is the collection of observed values from $\mathbf{x}_{i}$, and $\mathcal{A}_{i}$ is the set of configurations of $\mathbf{x}_{ic}$ that is consistent with the values in $\mathbf{x}_{i,obs}$.

Letting $\tilde{\mathbf{v}} = \{(y_{1}, W_{1}, \mathbf{x}_{1c}), \ldots, (y_{n}, W_{n}, \mathbf{x}_{nc})\}$ denote the ``complete data", the  complete-data log-likelihood associated with (\ref{eq:full_hierarchy}) is
\begin{eqnarray}
\ell_{c}(\bbeta, \bgamma|\tilde{\mathbf{v}})
= C + \sum_{i=1}^{n} u_{i}\mathbf{x}_{ic}^{T}\bbeta - \frac{1}{2}\sum_{i=1}^{n} W_{i}(\mathbf{x}_{ic}^{T}\bbeta)^{2}  +
\sum_{i=1}^{n} \sum_{k=1}^{2^{p-1}}I_{k}(\mathbf{x}_{i}) \log( \gamma_{k} ), \nonumber 
\end{eqnarray}
where $C$ is a term that does not depend on $(\bbeta, \bgamma)$.
The ``Q-function" associated with $\ell_{c}(\bbeta, \bgamma|\tilde{\mathbf{v}})$ is then
\begin{eqnarray}
Q(\bbeta, \bgamma|\bbeta^{(t)}, \bgamma^{(t)}) &=& E\{ \ell_{c}(\bbeta, \bgamma|\tilde{\mathbf{v}} ) \mid \mathbf{y}, \bbeta^{(t)}, \bgamma^{(t)} \} \nonumber \\
&=& C + \sum_{i=1}^{n} u_{i}E\{ \mathbf{x}_{ic}^{T} \mid \mathbf{y}, \bbeta^{(t)}, \bgamma^{(t)} \}\bbeta \nonumber \\
&-& \frac{1}{2}\sum_{i=1}^{n} E\big\{ W_{i}\bbeta^{T}\mathbf{x}_{ic}\mathbf{x}_{ic}^{T}\bbeta \mid \mathbf{y}, \bbeta^{(t)}, \bgamma^{(t)}\} 
+  \sum_{k=1}^{2^{p-1}} \log(\gamma_{k}) \sum_{i=1}^{n} I_{k}(\mathbf{x}_{ic}) \nonumber \\
&=& C + \sum_{i=1}^{n} u_{i}(\mathbf{a}_{i}^{(t)})^{T}\bbeta - \frac{1}{2}\sum_{i=1}^{n} \bbeta^{T}\mathbf{B}_{i}^{(t)}\bbeta +
\sum_{k=1}^{2^{p-1}} G_{k.}^{(t)}\log( \gamma_{k} ), 
\label{eq:qfn_miss}
\end{eqnarray}
where $\mathbf{a}_{i}^{(t)}$, $\mathbf{B}_{i}^{(t)}$, and $G_{k.}^{(t)}$ are defined as 
\begin{eqnarray}
\mathbf{a}_{i}^{(t)} &=& E\{ \mathbf{x}_{ic} \mid \mathbf{y}, \mathbf{x}_{i,obs}, \bbeta^{(t)}, \bgamma^{(t)} \} \nonumber \\ 
\mathbf{B}_{i}^{(t)} &=& E\{ W_{i}\mathbf{x}_{ic}\mathbf{x}_{ic}^{T} \mid \mathbf{y}, \mathbf{x}_{i,obs}, \bbeta^{(t)}, \bgamma^{(t)} \} \nonumber \\
G_{k.}^{(t)} &=& \sum_{i=1}^{n} E\{ I_{k}(\mathbf{x}_{i}) \mid \mathbf{y}, \mathbf{x}_{i,obs}, \bbeta^{(t)}, \bgamma^{(t)} \}. \nonumber 
\end{eqnarray}
Thus, we can re-write the Q-function (\ref{eq:qfn_miss}) as
\begin{equation}
Q(\bbeta, \bgamma|\bbeta^{(t)}, \bgamma^{(t)})
= C + \bbeta^{T}(\mathbf{A}^{(t)})^{T}\mathbf{u} - \frac{1}{2}\bbeta^{T}\mathbf{B}^{(t)}\bbeta  + \sum_{k=1}^{2^{p-1}} G_{k.}^{(t)}\log( \gamma_{k} ),
\label{eq:qfn_simple}
\end{equation}
where $\mathbf{A}^{(t)}$ is the $n \times p$ matrix whose $i^{th}$ row is $\mathbf{a}_{i}^{(t)}$ and $\mathbf{B}^{(t)}$ is the $p \times p$ matrix 
defined as
\begin{equation}
\mathbf{B}^{(t)} = \sum_{i=1}^{n} E\{W_{i}\mathbf{x}_{ic}\mathbf{x}_{ic}^{T}|\mathbf{y}, \bbeta^{(t)}, \bgamma^{(t)} \}. \nonumber 
\end{equation}
Maximizing (\ref{eq:qfn_simple}) subject to the constraint that $\gamma_{k} \geq 0$ and $\sum_{k=1}^{2^{p-1}} \gamma_{k} = 1$ yields the following parameter updates:
\begin{eqnarray}
\bbeta^{(t+1),EM} &=& \big[ \mathbf{B}^{(t)} \big]^{-1}(\mathbf{A}^{(t)})^{T}\mathbf{u} 
\nonumber \\
\gamma_{k}^{(t + 1),EM} &=& \frac{G_{k.}^{(t)}}{ \sum_{k=1}^{2^{p-1}} G_{k.}^{(t)}} \nonumber 
\end{eqnarray}

\bigskip

\noindent
To complete the description of this EM algorithm, we just need to describe
how to compute $\mathbf{A}^{(t)}$ and $\mathbf{B}^{(t)}$. To this end,
note that 
\begin{equation}
\mathbf{a}_{i}^{(t)} = \sum_{k=1}^{2^{p}} p_{ik}^{(t)}\mathbf{d}_{k}, \nonumber
\end{equation}
where $\mathbf{d}_{k}$ is one of the 
$2^{p}$ possible values of $\mathbf{x}_{ic}$ and $p_{ik}^{(t)}$ is defined as 
\begin{equation}
p_{ik}^{(t)} = P(\mathbf{x}_{ic} = \mathbf{d}_{k}|y_{i}, \mathbf{x}_{i,obs}, \bbeta^{(t)}, \bgamma^{(t)})
= \frac{p(y_{i}|\mathbf{x}_{ic}=\mathbf{d}_{k}, \bbeta^{(t)})\gamma_{k}^{(t)}I(k \in \mathcal{A}_{i})}{
\sum_{h \in \mathcal{A}_{i}}p(y_{i}|\mathbf{x}_{ic}=\mathbf{d}_{h}, \bbeta^{(t)})\gamma_{h}^{(t)} }.
\label{eq:post_probs_formula}
\end{equation}
In (\ref{eq:post_probs_formula}), $\mathcal{A}_{i}$ is the set of configurations of $\mathbf{x}_{ic}$ that is consistent with the observed covariate vector $\mathbf{x}_{i,obs}$. Regarding $\mathbf{B}^{(t)}$, one can note that 
\begin{equation}
\mathbf{B}^{(t)} = \sum_{i=1}^{n} \omega(\mathbf{x}_{i}^{T}\bbeta^{(t)})E\{\mathbf{x}_{ic}\mathbf{x}_{ic}^{T} \mid \mathbf{y}, \bbeta^{(t)}, \bgamma^{(t)} \}
= \sum_{i=1}^{n} \omega(\mathbf{x}_{i}^{T}\bbeta^{(t)}) \sum_{k=1}^{2^{p}} p_{ik}^{(t)} \mathbf{d}_{k}\mathbf{d}_{k}^{T}, \nonumber 
\end{equation}
where $p_{ik}^{(t)}$ is as defined in (\ref{eq:post_probs_formula}). 

One option for a PX-ECME version of this EM algorithm is to define $\gamma_{k}^{(t+1)} = \gamma_{k}^{(t+1), EM}$ and $\bbeta^{(t+1)} = \rho^{(t+1)}\bbeta^{(t+1), EM}$ where $\rho^{(t+1)}$ is chosen to maximize
the following ``observed" log-likelihood
\begin{eqnarray}
\ell_{o}(\rho\bbeta^{(t+1),EM}) 
&=& \sum_{i=1}^{n} \log\Big( p(y_{i}|\mathbf{x}_{i,obs}) \Big) \nonumber \\
&=& \sum_{i=1}^{n} \log\Bigg( \sum_{k=1}^{2^{p}}  p(y_{i}|\mathbf{x}_{ic}=\mathbf{d}_{k}, \bbeta=\rho\bbeta^{(t+1),EM})p(\mathbf{x}_{ic}=\mathbf{d}_{k}|\mathbf{x}_{i,obs}, \bgamma^{(t+1),EM}) I(k \in \mathcal{A}_{i}) \Bigg). \nonumber 
\end{eqnarray}

\section{Proof of Proposition 1 and Theorem 1}

\noindent
\textbf{Proof of Proposition 1.} 
First, note that 
\begin{eqnarray}
&& Q_{PX}^{\boldsymbol{\eta}}(\bbeta^{(t+1), GEM}, 1|\bbeta^{(t)}, \alpha_{0}) - Q_{PX}^{\boldsymbol{\eta}}(\bbeta^{(t)}, 1|\bbeta^{(t)}, \alpha_{0}) \nonumber \\
&=& \tilde{Q}_{PX}^{\boldsymbol{\eta}}(\bbeta^{(t+1), GEM}, 1|\bbeta^{(t)}, \alpha_{0}) - \tilde{Q}_{PX}^{\boldsymbol{\eta}}(\bbeta^{(t)}, 1|\bbeta^{(t)}, \alpha_{0}) 
+ \frac{1}{2}(\bbeta^{(t+1), GEM} - \bbeta^{(t)})^{T}\mathbf{H}^{(t)}(\bbeta^{(t+1), GEM} - \bbeta^{(t)}) \nonumber \\
&\geq& \tilde{Q}_{PX}^{\boldsymbol{\eta}}(\bbeta^{(t+1), GEM}, 1|\bbeta^{(t)}, \alpha_{0}) - \tilde{Q}_{PX}^{\boldsymbol{\eta}}(\bbeta^{(t)}, 1|\bbeta^{(t)}, \alpha_{0}) \nonumber \\ 
&\geq& 0, \nonumber
\end{eqnarray}
where the first inequality follows from the fact that $\mathbf{H}^{(t)}$ is positive semi-definite and
the second inequality follows from the fact that $\bbeta^{(t+1), GEM}$ maximizes $\tilde{Q}_{PX}^{\boldsymbol{\eta}}(\bbeta, 1|\bbeta^{(t)}, \alpha_{0})$. It then follows from Jensen's inequality and the above inequality that
\begin{equation}
p\ell_{o, \mathbf{s}}^{\boldsymbol{\eta}}( \bbeta^{(t+1), GEM}|\mathbf{y}) -  p\ell_{o, \mathbf{s}}^{\boldsymbol{\eta}}( \bbeta^{(t)}|\mathbf{y})
\geq \tilde{Q}_{PX}^{\boldsymbol{\eta}}(\bbeta^{(t+1), GEM}, 1|\bbeta^{(t)}, \alpha_{0}) - \tilde{Q}_{PX}^{\boldsymbol{\eta}}(\bbeta^{(t)}, 1|\bbeta^{(t)}, \alpha_{0}) \geq 0 \nonumber
\end{equation}
Then, since $\rho^{(t+1)} = \argmax_{\rho \in \mathbb{R}} p\ell_{o, \mathbf{s}}^{\boldsymbol{\eta}}(\rho\bbeta^{(t+1), GEM}|\mathbf{y})$, we have that
\begin{eqnarray}
p\ell_{o, \mathbf{s}}^{\boldsymbol{\eta}}( \bbeta^{(t+1)}|\mathbf{y}) &=& p\ell_{o, \mathbf{s}}^{\boldsymbol{\eta}}( \rho^{(t+1)}\bbeta^{(t+1), GEM}|\mathbf{y}) \nonumber \\
&\geq& p\ell_{o, \mathbf{s}}^{\boldsymbol{\eta}}( \bbeta^{(t+1), GEM}|\mathbf{y}) \nonumber \\
&\geq& p\ell_{o, \mathbf{s}}^{\boldsymbol{\eta}}( \bbeta^{(t)}|\mathbf{y}). \nonumber 
\end{eqnarray}

\noindent
\textbf{Proof of Theorem 1.} First, recall that the PX-ECME mapping $G_{PX}$ can be expressed in terms of the EM mapping $G_{EM}$ as
\begin{equation}
G_{PX}( \bbeta ) = h( G_{EM}(\bbeta) )G_{EM}( \bbeta ) = a(\bbeta)G_{EM}(\bbeta), \nonumber
\end{equation}
where $h(\bbeta) = \argmax_{\rho \in \mathbb{R}} \ell_{o,\mathbf{s}}(\rho\bbeta|\mathbf{y})$ and $a(\bbeta) = h(G_{EM}(\bbeta))$. Hence, if $[G_{PX}(\bbeta)]_{j}$ denotes the $j^{th}$ component of $G_{PX}(\bbeta)$, then $\partial [G_{PX}(\bbeta)]_{j}/\partial \beta_{k} = a( G_{EM}(\bbeta) )\partial [G_{EM}(\bbeta)]_{j}/\partial \beta_{k} + [G_{EM}(\bbeta)]_{j}\partial a(\bbeta)/\partial \beta_{k}$. This implies that the Jacobian matrix $J_{PX}(\bbeta)$ of $G_{PX}(\bbeta)$ can be expressed as
\begin{equation}
J_{PX}( \bbeta ) = \frac{\partial G_{PX}(\bbeta)}{\partial \bbeta} = a(\bbeta)J_{EM}(\bbeta) + G_{EM}(\bbeta)\nabla a(\bbeta)^{T},
\label{eq:jacobian_decomposition}
\end{equation}
where $\nabla a(\bbeta)$ denotes the gradient of $a(\bbeta)$ and $J_{EM}(\bbeta)$ is the Jacobian matrix associated with the EM mapping $G_{EM}(\bbeta)$.

Now, note that we can express the gradient of $a(\bbeta)$ in terms
of the gradient of $h(\bbeta)$ evaluated at $G_{EM}(\bbeta)$ as
\begin{equation}
\nabla a(\bbeta)^{T} = \nabla h( G_{EM}(\bbeta) )^{T}J_{EM}(\bbeta)
\end{equation}

To derive an explicit formula for $\nabla h(\bbeta)$, we first note that $h(\bbeta)$ is defined implicitly by the equation $s(\bbeta, h(\bbeta)) = 0$ where
the score function $s(\bbeta, h(\bbeta))$ is defined as
\begin{eqnarray}
s(\bbeta, h(\bbeta)) &=& \sum_{i=1}^{n} s_{i}y_{i}\mathbf{x}_{i}^{T}\bbeta
- \sum_{i=1}^{n} s_{i}m_{i}\mathbf{x}_{i}^{T}\bbeta\pi\{ h(\bbeta)\mathbf{x}_{i}^{T}\bbeta \} \nonumber \\
&=& \bbeta^{T}\mathbf{X}^{T}\mathbf{S}\mathbf{y} - \bbeta^{T}\mathbf{X}\mathbf{S}\mu\{ h(\bbeta)\mathbf{X}\bbeta \}, \nonumber
\end{eqnarray}
where $\mu\{ h(\bbeta)\mathbf{X}\bbeta \}$ is the length-n vector whose $i^{th}$ element is $m_{i}\pi\{ h(\bbeta)\mathbf{x}_{i}^{T}\bbeta\}$.
If we differentiate $s(\bbeta, h(\bbeta))$ with respect to $h(\bbeta)$, we obtain 
\begin{eqnarray}
\frac{\partial s(\bbeta, h(\bbeta))}{\partial h}
&=& - \sum_{i=1}^{n} m_{i}s_{i}(\mathbf{x}_{i}^{T}\bbeta)^{2}
\pi\{ h(\bbeta)\mathbf{x}_{i}^{T}\bbeta \}\big[ 1 - \pi\{ h(\bbeta)\mathbf{x}_{i}^{T}\bbeta \} \big] \nonumber \\
&=& \bbeta^{T}\mathbf{X}^{T}\mathbf{S}\mathbf{R}(\bbeta)\mathbf{X}\bbeta \nonumber \\
&=& c(\bbeta), \nonumber 
\end{eqnarray}
where $R(\bbeta)$ is the diagonal matrix whose $i^{th}$ diagonal element is
$ m_{i}\pi\{h(\bbeta)\mathbf{x}_{i}^{T}\bbeta\}[1 - \pi\{h(\bbeta)\mathbf{x}_{i}^{T}\bbeta\}]$
and $\pi(u) = \{ 1 + \exp(-u) \}^{-1}$. 
Now, if we differentiate $s(\bbeta, h(\bbeta))$ with respect to $\beta_{k}$ (keeping $h(\bbeta)$ fixed), we obtain 
\begin{eqnarray}
\frac{\partial s(\bbeta, h(\bbeta))}{\partial \beta_{k}}
&=& \sum_{i=1}^{n} s_{i}y_{i}x_{ik} - \sum_{i=1}^{n} s_{i}m_{i}x_{ik}\pi\{ h(\bbeta)\mathbf{x}_{i}^{T}\bbeta \}
\nonumber \\
&-& h(\bbeta)\sum_{i=1}^{n} s_{i}m_{i}x_{ik}\mathbf{x}_{i}^{T}\bbeta\pi\{ h(\bbeta)\mathbf{x}_{i}^{T}\bbeta \}
[1 - \pi\{ h(\bbeta)\mathbf{x}_{i}^{T}\bbeta \}]. \nonumber 
\end{eqnarray}
So, we can write the $p \times 1$ fixed-h gradient $\nabla_{\beta} s(\beta, h(\bbeta)) = \big(\tfrac{\partial s(\boldsymbol{\beta}, h(\boldsymbol{\beta}))}{\partial \beta_{1}} , ...., \tfrac{\partial s(\boldsymbol{\beta}, h(\boldsymbol{\beta}))}{\partial \beta_{p}} \big)^{T}$
as 
\begin{equation}
\nabla_{\beta} s(\bbeta, h(\bbeta))
= \mathbf{X}^{T}\mathbf{S}[ \mathbf{y} - \mu\{ h(\bbeta)\mathbf{X}\bbeta \}] - h(\bbeta)\mathbf{X}^{T}\mathbf{S}\mathbf{R}(\bbeta)\mathbf{X}\bbeta. \nonumber
\end{equation}
By the implicit function theorem for continuously differentiable functions, we can express the gradient $\nabla h(\bbeta)$ as
\begin{eqnarray}
\nabla h(\bbeta) &=& -\Big[ \frac{s(\bbeta, h(\bbeta))}{\partial h}\Big]^{-1} \nabla_{\beta} s(\bbeta, h(\bbeta)) \nonumber \\
&=& \frac{1}{c(\bbeta)}\Big[ \mathbf{X}^{T}\mathbf{S}[ \mathbf{y} - \mu\{ h(\bbeta)\mathbf{X}\bbeta \}] - h(\bbeta)\mathbf{X}^{T}\mathbf{S}\mathbf{R}(\bbeta)\mathbf{X}\bbeta \Big]. \nonumber 
\end{eqnarray}
At the point of convergence $\bbeta^{*} = G_{EM}(\bbeta^{*})$, $h(\bbeta^{*}) = 1$ and
$\mathbf{X}^{T}\mathbf{S}[ \mathbf{y} - \mu\{ h(\bbeta^{*})\mathbf{X}\bbeta^{*} \}] = \mathbf{0}$ and hence
\begin{equation}
\nabla h( G_{EM}(\bbeta^{*}) ) = \frac{-1}{c(\bbeta^{*})}\Big[ \mathbf{X}^{T}\mathbf{S}\mathbf{R}(\bbeta^{*})\mathbf{X}\bbeta^{*} \Big]. \nonumber 
\end{equation}
Finally, returning to (\ref{eq:jacobian_decomposition}), we have that
\begin{eqnarray}
J_{PX}(\bbeta^{*}) &=& a(\bbeta^{*})J_{EM}(\bbeta^{*}) + G_{EM}(\bbeta^{*})\nabla a(\bbeta^{*})^{T} \nonumber \\
&=& a(\bbeta^{*})J_{EM}(\bbeta^{*}) + G_{EM}(\bbeta^{*})\nabla h(G_{EM}(\bbeta^{*}))^{T} J_{EM}(\bbeta^{*}) \nonumber \\
&=& J_{EM}(\bbeta^{*}) + \bbeta^{*}\nabla h(G_{EM}(\bbeta^{*}))^{T}J_{EM}(\bbeta^{*}) \nonumber \\
&=& J_{EM}(\bbeta^{*}) - \frac{1}{c(\bbeta^{*})}\bbeta^{*}(\bbeta^{*})^{T} \mathbf{X}^{T}\mathbf{S}\mathbf{R}(\bbeta^{*})\mathbf{X}J_{EM}(\bbeta^{*}) \nonumber \\
&=& J_{EM}(\bbeta^{*}) - \frac{1}{c(\bbeta^{*})}\bbeta^{*}(\bbeta^{*})^{T} \mathbf{X}^{T}\mathbf{S}\mathbf{R}(\bbeta^{*})\mathbf{X}\Big[ \mathbf{I}_{p}  - (\mathbf{X}^{T}\mathbf{S}\mathbf{W}(\bbeta^{*})\mathbf{X})^{-1} \mathbf{X}^{T}\mathbf{S}\mathbf{R}(\bbeta^{*})\mathbf{X}\Big] \nonumber \\
&=& J_{EM}(\bbeta^{*}) - \frac{1}{c(\bbeta^{*})}\bbeta^{*}(\bbeta^{*})^{T} \mathbf{V}(\bbeta^{*})\mathbf{A}(\bbeta^{*})\mathbf{V}(\bbeta^{*})^{T}, \nonumber 
\end{eqnarray}
where $\mathbf{V}(\bbeta^{*})$ and $\mathbf{A}(\bbeta^{*})$ are defined as
\begin{eqnarray}
\mathbf{V}(\bbeta^{*}) &=& 
\mathbf{X}^{T}\mathbf{S}\mathbf{R}(\bbeta^{*})\mathbf{X} \nonumber \\
\mathbf{A}(\bbeta^{*}) &=& (\mathbf{X}^{T}\mathbf{S}\mathbf{R}(\bbeta^{*})\mathbf{X})^{-1} - (\mathbf{X}^{T}\mathbf{S}\mathbf{W}(\bbeta^{*})\mathbf{X})^{-1}. \nonumber 
\end{eqnarray}


\noindent
We now turn to the question of comparing the spectral radii of $\mathbf{J}_{PX}(\bbeta^{*})$ and $\mathbf{J}_{EM}(\bbeta^{*})$.
To address this, we first consider the matrices $\mathbf{S}_{PX}$ and $\mathbf{S}_{EM}$ defined as
\begin{eqnarray}
\mathbf{S}_{PX} &=& \mathbf{I} - \mathbf{J}_{PX}(\bbeta^{*}) = \mathbf{I} - \mathbf{J}_{EM}(\bbeta^{*}) + [c(\bbeta^{*})]^{-1}\bbeta^{*}(\bbeta^{*})^{T}\mathbf{V}(\bbeta^{*})\mathbf{A}(\bbeta^{*})\mathbf{V}(\bbeta^{*})^{T} \nonumber \\
\mathbf{S}_{EM} &=& \mathbf{I} - \mathbf{J}_{EM}(\bbeta^{*}). \nonumber 
\end{eqnarray}
Because the diagonal elements of $\mathbf{W}(\bbeta^{*})$ are greater than the corresponding diagonal elements of $\mathbf{R}(\bbeta^{*})$ (i.e., $\pi(\mathbf{x}_{i}^{T}\bbeta)\{ 1 - \pi(\mathbf{x}_{i}^{T}\bbeta) \} \leq \tanh(\mathbf{x}_{i}^{T}\bbeta/2)/2\mathbf{x}_{i}^{T}\bbeta$),
$\mathbf{A}(\bbeta^{*})$ is a symmetric positive definite matrix.
This implies that $\mathbf{V}(\bbeta^{*})\mathbf{A}(\bbeta^{*})\mathbf{V}(\bbeta^{*})^{T}$ is also a symmetric positive definite matrix. 
Hence, $\bbeta^{*}(\bbeta^{*})^{T}\mathbf{V}(\bbeta^{*})\mathbf{A}(\bbeta^{*})\mathbf{V}(\bbeta^{*})^{T}$ is symmetric positive definite
as it is the product of two symmetric positive definite matrices.
Hence, because $c(\bbeta^{*}) > 0$, we have $\mathbf{S}_{PX} \succeq \mathbf{S}_{EM}$, which then implies that the $r_{PX} \leq r_{EM}$, where $r_{PX}$ and $r_{EM}$ are the largest eigenvalues of $J_{PX}(\bbeta^{*})$ and $J_{EM}(\bbeta^{*})$ respectively.

\section{PX-ECME and Coordinate Descent}

\begin{algorithm}[H]
\setstretch{1.25}
\caption{(PX-ECME Coordinate Descent for Logistic Regression with Elastic Net Penalty). 
In the description of the algorithm,
$\mathbf{u} = (u_{1}, \ldots, u_{n})^{T}$, where $u_{i} = y_{i} - m_{i}/2$.}
\begin{algorithmic}[1]
\State Given an integer $1 \leq k \leq p$, $\bbeta^{(0)} \in \mathbb{R}^{p}$ and $\alpha^{(0)} \in \mathbb{R}$. Set $\btheta^{(0)} = \alpha^{(0)}\bbeta^{(0)}$.
\State Set $\mathbf{W}(\bbeta^{(0)}) = \textrm{diag}\{ m_{1}/4, \ldots, m_{n}/4\}$.
\For{t=0,1,2,... until convergence}
\For{j = 1,..., p} 

\hspace{0.3cm} Update the components $\theta_{j}^{(t+1)}$ of $\btheta^{(t+j/p)}$:
\vspace{-.2cm}
\begin{eqnarray}
\theta_{j}^{(t+1)} &=& \textrm{sign}\Big( \tfrac{1}{ \alpha^{(t + j/p)} } V_{j}(\bbeta^{(t + j/p)}) - U_{j}(\bbeta^{(t + j/p)}, \btheta^{(t + j/p)})  \Big) \nonumber \\
&\times& \max\Big\{ \Big| \tfrac{1}{ \alpha^{(t + j/p)} }V_{j}(\bbeta^{(t + j/p)}) - U_{j}(\bbeta^{(t + j/p)}, \btheta^{(t + j/p)}) \Big| - \tfrac{1}{ \alpha^{(t + j/p)} }\tilde{\lambda}(\bbeta^{(t + j/p)}), 0 \Big\}, \nonumber
\end{eqnarray}
\hspace{1.1cm} where $V_{j}(\bbeta^{(t + j/p)})$, $U_{j}(\bbeta^{(t + j/p)}, \btheta_{-j}^{(t + j/p)})$, and $\tilde{\lambda}(\bbeta^{(t + j/p)})$ are as defined in (\ref{eq:vuterms}).
\If{ $j$ mod $k = 0$ \textbf{or} $j = p$} 
\State Compute
\begin{equation}
\alpha^{(t+j/p)} = \argmax_{\alpha \in \mathbb{R}}\Big[ p\ell_{o,\mathbf{s}}^{\boldsymbol{\eta}}( \alpha \btheta^{(t+1)}|\mathbf{y}) \Big]. \nonumber
\end{equation}
\State Set $\bbeta^{(t+j/p)} = \alpha^{(t+j/p)}\btheta^{(t+j/p)}$.
\State Update the diagonals $\omega(\mathbf{x}_{i}^{T}\bbeta^{(t + j/p)}, m_{i})$  of the weight matrix $\mathbf{W}(\bbeta^{(t + j/p)})$
\begin{equation}
\omega(\mathbf{x}_{i}^{T}\bbeta^{(t + j/p)}, m_{i}) = \frac{m_{i}}{2\mathbf{x}_{i}^{T}\bbeta^{(t + j/p)}}\tanh\big( \mathbf{x}_{i}^{T}\bbeta^{(t + j/p)}/2 \big). \nonumber
\end{equation}
\Else
\State Set $\bbeta^{(t + j/p)} = \bbeta^{t + (j-1)/p}$, $\alpha^{(t + j/p)} = \alpha^{t + (j-1)/p}$,  $\mathbf{W}(\bbeta^{(t + j/p)}) = \mathbf{W}(\bbeta^{(t + (j-1)/p})$.
\EndIf
\EndFor
\EndFor
\end{algorithmic}
\label{alg:pxecme_coord_desc}
\end{algorithm}

The terms $V_{j}(\bbeta^{(t)})$, $U_{j}(\bbeta^{(t)}, \btheta_{-j}^{(t + j/p)})$, and $\tilde{\lambda}(\bbeta^{(t)})$ mentioned in Section \ref{sec:coord_descent} are given by
\begin{eqnarray}
V_{j}(\bbeta^{(t)}) &=& \sum_{i=1}^{n} x_{ij}s_{i}u_{i} \big/\Big\{ \sum_{i=1}^{n}(A_{ij}^{(t)})^{2} + \lambda_{2} \Big\} \nonumber \\
U_{j}(\bbeta^{(t)}, \btheta_{-j}^{(t + j/p)}) &=& \sum_{i=1}^{n} \Big\{ A_{ij}^{(t)}\sum_{k \neq j}\theta_{k}^{(t + j/p)}A_{ik}^{(t)} \Big\}\Big/ \Big\{ \sum_{i=1}^{n}(A_{ij}^{(t)})^{2} + \lambda_{2} \Big\} \nonumber \\
\tilde{\lambda}(\bbeta^{(t)}) &=& \lambda_{1}\big/\Big\{ \sum_{i=1}^{n}(A_{ij}^{(t)})^{2} + \lambda_{2} \Big\}. \label{eq:vuterms}
\end{eqnarray}

\bibliographystyle{agsm}
\bibliography{PXECME_refs}

@article{bohning1988,
  title={Monotonicity of quadratic-approximation algorithms},
  author={B{\"o}hning, Dankmar and Lindsay, Bruce G},
  journal={Annals of the Institute of Statistical Mathematics},
  volume={40},
  number={4},
  pages={641--663},
  year={1988}
}

@book{brent2013,
  title={Algorithms for minimization without derivatives},
  author={Brent, Richard P},
  year={2013},
  publisher={Courier Corporation}
}

@book{chambers1992,
  title={Statistical models in S},
  author={Chambers, John M and Hastie, Trevor J},
  year={1992},
  publisher={Wadsworth and Brooks/Cole},
  address = {Pacific Grove, CA}
}

@article{choi2013,
  title={The {P}{\'o}lya--{G}amma {G}ibbs sampler for {B}ayesian logistic regression is uniformly ergodic},
  author={Choi, Hee Min and Hobert, James P and others},
  journal={Electronic Journal of Statistics},
  volume={7},
  pages={2054--2064},
  year={2013}
}

@article{durante2018,
  title={A note on quadratic approximations of logistic log-likelihoods},
  author={Durante, Daniele and Rigon, Tommaso},
  journal={ArXiv},
  volume={1711},
  year={2018}
}

@article{fan2001,
  title={Variable selection via nonconcave penalized likelihood and its oracle properties},
  author={Fan, Jianqing and Li, Runze},
  journal={Journal of the American statistical Association},
  volume={96},
  number={456},
  pages={1348--1360},
  year={2001}
}

@article{Friedman2007,
  title={Pathwise coordinate optimization},
  author={Friedman, Jerome and Hastie, Trevor and H{\"o}fling, Holger and Tibshirani, Robert and others},
  journal={The annals of applied statistics},
  volume={1},
  number={2},
  pages={302--332},
  year={2007}
}

@article{Friedman2010,
  title={Regularization paths for generalized linear models via coordinate descent},
  author={Friedman, Jerome and Hastie, Trevor and Tibshirani, Rob},
  journal={Journal of statistical software},
  volume={33},
  number={1},
  pages={1},
  year={2010}
}

@article{Green1984,
  title={Iteratively reweighted least squares for maximum likelihood estimation, and some robust and resistant alternatives},
  author={Green, Peter J},
  journal={Journal of the Royal Statistical Society: Series B},
  volume={46},
  number={2},
  pages={149--170},
  year={1984}
}

@article{guyon2004,
  title={Result analysis of the {NIPS} 2003 feature selection challenge},
  author={Guyon, Isabelle and Gunn, Steve and Ben-Hur, Asa and Dror, Gideon},
  journal={Advances in neural information processing systems},
  volume={17},
  year={2004}
}

@article{hunter2004,
  title={A tutorial on {MM} algorithms},
  author={Hunter, David R and Lange, Kenneth},
  journal={The American Statistician},
  volume={58},
  number={1},
  pages={30--37},
  year={2004}
}

@article{ibrahim1990,
  title={Incomplete data in generalized linear models},
  author={Ibrahim, Joseph G},
  journal={Journal of the American Statistical Association},
  volume={85},
  number={411},
  pages={765--769},
  year={1990}
}

@article{ibrahim1999,
  title={Monte {C}arlo {EM} for missing covariates in parametric regression models},
  author={Ibrahim, Joseph G and Chen, Ming-Hui and Lipsitz, Stuart R},
  journal={Biometrics},
  volume={55},
  number={2},
  pages={591--596},
  year={1999}
}

@article{jaakkola2000,
  title={Bayesian parameter estimation via variational methods},
  author={Jaakkola, Tommi S and Jordan, Michael I},
  journal={Statistics and Computing},
  volume={10},
  pages={25--37},
  year={2000}
}

@book{Lange2012,
 author = {Lange, Kenneth},
 title = {Numerical Analysis for Statisticians},
 year = {2012},
 isbn = {146142612X, 9781461426127},
 edition = {2nd},
 publisher = {Springer Publishing Company, Incorporated}
}

@article{Lewandowski2010,
  title={Parameter expansion and efficient inference},
  author={Lewandowski, Andrew and Liu, Chuanhai and Wiel, Scott Vander},
  journal={Statistical Science},
  pages={533--544},
  year={2010}
}

@article{Liu1994,
 author = {Chuanhai Liu and Donald B. Rubin},
 journal = {Biometrika},
 number = {4},
 pages = {633-648},
 title = {The {ECME} algorithm: a simple extension of {EM} and {ECM} with faster monotone convergence},
 volume = {81},
 year = {1994}
}

@article{Liu1997,
  title={{ML} Estimation of the Multivariate t Distribution and the {EM} Algorithm},
  author={Liu, Chuanhai},
  journal={Journal of Multivariate Analysis},
  volume={63},
  number={2},
  pages={296--312},
  year={1997}
}

@article{Liu1998,
 author = {Chuanhai Liu and Donald B. Rubin and Ying Nian Wu},
 journal = {Biometrika},
 number = {4},
 pages = {755-770},
 title = {Parameter expansion to accelerate {EM}: the {PX-EM} algorithm},
 volume = {85},
 year = {1998}
}

@article{marschner2011,
  title={glm2: Fitting generalized linear models with convergence problems},
  author={Marschner, Ian C},
  journal={The R Journal},
  volume={3},
  number={2},
  pages={12--15},
  year={2011}
}

@book{mclachlan2007,
  title={The EM algorithm and extensions},
  author={McLachlan, Geoffrey J and Krishnan, Thriyambakam},
  volume={382},
  year={2007},
  publisher={John Wiley \& Sons}
}

@article{Meng1993,
 author = {Xiao Li Meng and Donald B. Rubin},
 title = {Maximum likelihood estimation via the {ECM} algorithm: a general framework},
 number = {2},
 pages = {267-278},
 journal = {Biometrika},
 volume = {80},
 year = {1993}
}

@article{Parikh2014,
  title={Proximal algorithms},
  author={Parikh, Neal and Boyd, Stephen},
  journal={Foundations and Trends in optimization},
  volume={1},
  number={3},
  pages={127--239},
  year={2014}
}

@article{Polson2013,
  title={Bayesian inference for logistic models using {P}{\'o}lya--{G}amma latent variables},
  author={Polson, Nicholas G and Scott, James G and Windle, Jesse},
  journal={Journal of the American Statistical Association},
  volume={108},
  number={504},
  pages={1339--1349},
  year={2013}
}

@article{Scott2013,
  title={Expectation-maximization for logistic regression},
  author={Scott, James G and Sun, Liang},
  journal={arXiv preprint arXiv:1306.0040},
  year={2013}
}

@article{vandyk2000,
  title={Fitting mixed-effects models using efficient {EM}-type algorithms},
  author={Van Dyk, David A},
  journal={Journal of Computational and Graphical Statistics},
  volume={9},
  number={1},
  pages={78--98},
  year={2000}
}

@article{Varadhan2008,
	Author = {Ravi Varadhan and Christophe Roland},
	Journal = {Scandinavian Journal of Statistics},
	Pages = {335--353},
	Title = {Simple and Globally convergent methods for accelerating the convergence of any {EM} algorithm},
	Volume = {35},
    Number = {2},
	Year = {2008}
}

@article{Walker2011,
	Author = {Homer F. Walker and Peng Ni},
	Journal = {SIAM Journal on Numerical Analysis},
	Pages = {1715--1735},
	Title = {Anderson acceleration for fixed-point iterations},
	Volume = {49},
    Number = {4},
	Year = {2011}
 }

@article{Zhou2011,
	Author = {Hua Zhou and David Alexander and Kenneth Lange},
	Journal = {Statistics and Computing},
	Pages = {261--273},
	Title = {A quasi-{N}ewton accleration for high-dimensional optimization algorithms},
	Volume = {21},
    Number = {2},
	Year = {2011}
 }

\end{document}